\documentclass[twocolumn,showpacs,preprintnumbers,superscriptaddress]{revtex4}%
\usepackage{amssymb}
\usepackage{amsmath}
\usepackage{graphicx}
\usepackage{dcolumn}
\usepackage{bm}
\usepackage{amsfonts}%
\usepackage{epstopdf}

\providecommand{\U}[1]{\protect\rule{.1in}{.1in}}
\providecommand{\U}[1]{\protect\rule{.1in}{.1in}}
\begin{document}
\title{Superharmonic Resonances in a Strongly Coupled Cavity-Atom System}
\author{Eyal Buks}
\affiliation{Department of Electrical Engineering, Technion, Haifa 32000 Israel}
\author{Chunqing Deng}
\affiliation{Institute for Quantum Computing, University of Waterloo, Waterloo, ON, Canada N2L 3G1}
\affiliation{Department of Physics and Astronomy, University of Waterloo, Waterloo, ON, Canada N2L 3G1}
\affiliation{Waterloo Institute for Nanotechnology, University of Waterloo, Waterloo, ON, Canada N2L 3G1}
\author{Jean-Luc F.X. Orgazzi}
\affiliation{Institute for Quantum Computing, University of Waterloo, Waterloo, ON, Canada N2L 3G1}
\affiliation{Department of Electrical and Computer Engineering, University of Waterloo, Waterloo, ON, Canada N2L 3G1}
\affiliation{Waterloo Institute for Nanotechnology, University of Waterloo, Waterloo, ON, Canada N2L 3G1}
\author{Martin Otto}
\affiliation{Institute for Quantum Computing, University of Waterloo, Waterloo, ON, Canada N2L 3G1}
\affiliation{Department of Physics and Astronomy, University of Waterloo, Waterloo, ON, Canada N2L 3G1}
\affiliation{Waterloo Institute for Nanotechnology, University of Waterloo, Waterloo, ON, Canada N2L 3G1}
\author{Adrian Lupascu}
\affiliation{Institute for Quantum Computing, University of Waterloo, Waterloo, ON, Canada N2L 3G1}
\affiliation{Department of Physics and Astronomy, University of Waterloo, Waterloo, ON, Canada N2L 3G1}
\affiliation{Waterloo Institute for Nanotechnology, University of Waterloo, Waterloo, ON, Canada N2L 3G1}
\date{\today }

\begin{abstract}
We study a system consisting of a superconducting flux qubit strongly coupled
to a microwave cavity. The fundamental cavity mode is externally driven and the response is
investigated in the weak nonlinear regime. We find that near the crossing
point, at which the resonance frequencies of the cavity mode and qubit
coincide, the sign of the Kerr coefficient changes, and consequently the type
of nonlinear response changes from softening to hardening. Furthermore, the
cavity response exhibits superharmonic resonances when the ratio between the
qubit frequency and the cavity fundamental mode frequency is tuned close to an integer
value. The nonlinear response is characterized by the method of
intermodulation and both signal and idler gains are measured. The experimental
results are compared with theoretical predictions and good qualitative agreement is
obtained. The superharmonic resonances have potential for applications in
quantum amplification and generation of entangled states of light.

\end{abstract}
\maketitle





Cavity quantum electrodynamics (CQED) \cite{Haroche_24} is the study of the
interaction between photons confined in a cavity and atoms (natural or
artificial). The interaction is commonly described by the Rabi or
Jaynes-Cummings Hamiltonians \cite{Shore_1195}, and it has been the subject of
numerous theoretical and experimental investigations. An on-chip CQED system
can be realized by integrating a Josephson qubit
\cite{Clarke_1031,Devoret_1169,You_589} (playing the role of an artificial
atom) with a superconducting microwave resonator (cavity)
\cite{Blais_062320,Wallraff_162,Schoelkopf_664}. Superconducting CQED systems
have generated a fast growing interest due to the possibility to reach the
strong \cite{Wallraff_162}\ and ultra-strong \cite{Niemczyk_772,Forn_237001}
coupling regimes, and due to potential applications in quantum information
processing \cite{Devoret_1169,Barends_500,Chow_1,Riste_1}.

\begin{figure}
[ptb]
\begin{center}
\includegraphics[
height=2.60in,
width=2.9in
]%
{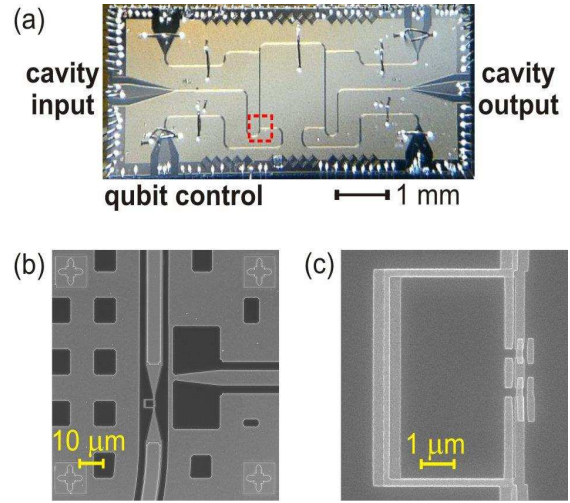}%
\caption{(color online) The CQED device. (a) Optical image of a device used in the current experiment, which is nominally
identical to the one described in \cite{Orgiazzi_1407_1346}, where the overlaid
dashed rectangle indicates the position of the qubit under study. The resonator, which occupies the central part of the chip, is coupled to the input/output ports using capacitors. (b) Electron micrograph
showing a qubit embedded in the coplanar waveguide resonator and its local
flux control line. (c) Electron micrograph of a flux qubit.}%
\label{Fig device}%
\end{center}
\end{figure}

In this study we investigate the driven dynamics of a strongly interacting
system composed of a superconducting flux qubit
\cite{Mooij_1036,Orlando_15398} and a coplanar waveguide (CPW) microwave
cavity
\cite{Niemczyk_772,Abdumalikov_180502,Bal_1324,Orgiazzi_1407_1346,Jerger_042604,Oelsner_172505,Inomata_140508}%
. The nonlinear cavity response
\cite{Serban_022305,Laflamme_033803,Siddiqi_207002,Lupacscu_127003,Boaknin_0702445,Mallet_791,Boissonneault_060305,Boissonneault_100504,Boissonneault_022324,Boissonneault_022305,Boissonneault_013819,Reed_173601,Ong_047001,Ong_167002,Bishop_105}
is measured as a function of the magnetic flux that is applied to the qubit.
At weak driving and when the ratio between the qubit frequency and the cavity
fundamental mode frequency is tuned close to the value $\omega_{\mathrm{a}}/\omega
_{\mathrm{c}}=1$ the common Jaynes-Cummings resonance, which henceforth is
referred to as the primary resonance, is observed. With stronger driving,
however, and when the ratio $\omega_{\mathrm{a}}/\omega_{\mathrm{c}}$ is tuned
close to integer values larger than unity, superharmonic resonances appear in
the measured response. Intermodulation (IMD) measurements are employed to characterize the nonlinear response 
\cite{Castellanos_083509,Vijay_110502,Yurke_5054}. The results are compared
with the predictions of a theoretical model, which is based on linearization
of the equations of motion that govern the dynamics of the CQED system under study.%

The investigated device contains a CPW cavity weakly coupled to two ports that
are used for performing microwave transmission measurements [see Fig.
\ref{Fig device}(a)]. Two persistent current flux qubits \cite{Mooij_1036},
consisting of a superconducting loop interrupted by four Josephson junctions
[see Fig. \ref{Fig device}(c)], are inductively coupled to the CPW resonator
[see Fig. \ref{Fig device}(b)]. In the current experiment, however, only one
qubit significantly affects the cavity mode response, whereas the other one is
made effectively decoupled by detuning its energy gap away from the mode
frequency. A CPW line terminated by a low inductance shunt is used to send
microwave pulses for coherent qubit control [see Fig. \ref{Fig device}(b)].
The device is fabricated on a high resistivity silicon substrate, in a two-step
process. In the first step, the resonator and the control lines are defined
using optical lithography, evaporation of a $190%
\operatorname{nm}%
$ thick aluminum layer and liftoff. In the second step, a bilayer resist is
patterned by electron-beam lithography. Subsequently, shadow evaporation of
two aluminum layers, $40%
\operatorname{nm}%
$ and $65%
\operatorname{nm}%
$ thick respectively, followed by liftoff define the qubit junctions.%

\begin{figure*}
[ptb]
\begin{center}
\includegraphics[
height=5.318in,
width=6.8807in
]%
{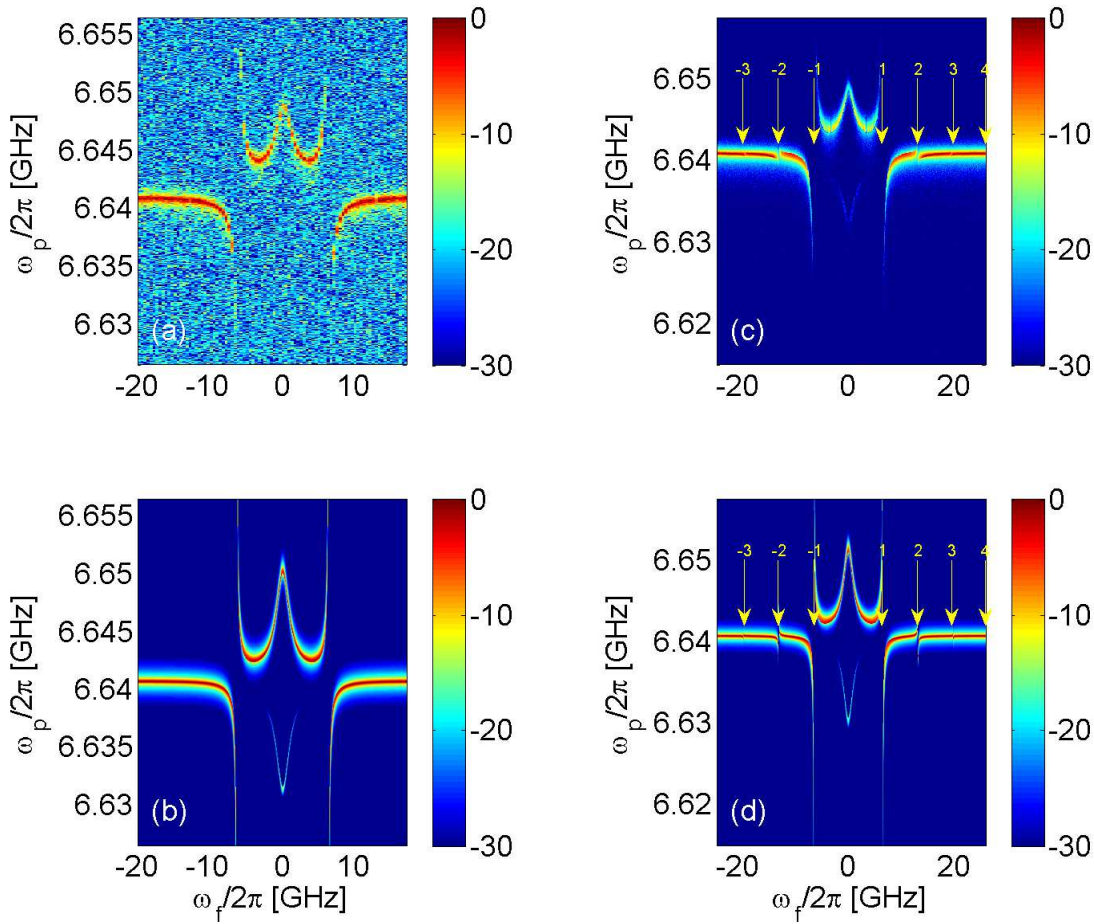}%
\caption{(color online) The measured (panels a and c) and calculated (panels b and d) cavity
transmission (in dB units) vs. $\omega_{\mathrm{f}}/2\pi$. For the panels on
the left (panels a and b) the power injected into the cavity is $-127$ dBm,
whereas for the panels on the right (panels c and d) the power is $-112$ dBm.
The following parameters have been assumed in the calculation:
$T=23\operatorname{mK}$, $\omega_{\mathrm{c}}/2\pi=6.6408\operatorname{GHz}%
$, $\omega_{\Delta}/2\pi=1.12\operatorname{GHz}$, $g/2\pi
=0.274\operatorname{GHz}$, $\gamma_{\mathrm{c}1}/\omega_{\mathrm{c}}%
=5\times10^{-6}$ and $\gamma_{\mathrm{c}2}=1.1\times\gamma_{\mathrm{c}1}$. The
relaxation time $T_{1}=1.2\operatorname{\mu s}\left(  1+0.45\operatorname{ns}%
\times\left\vert \omega_{\mathrm{f}}\right\vert \right)  $ is obtained from
energy relaxation measurements (in the range of qubit frequencies well below $\omega_{\mathrm{c}}/2\pi$), and the rate $T_{2}^{-1}=4.5\operatorname{MHz}%
\left(  1+44\left\vert \omega_{\mathrm{f}}\right\vert /\omega_{\mathrm{a}%
}\right)  $ is obtained from Ramsey rate measurements
\cite{Orgiazzi_1407_1346}. In panel a the measured off-resonance transmission
is significantly higher than the calculated one (see panel b) due to
instrumental noise, which has not been taken into account in the theoretical
modeling. In the region where $\Delta=\omega_{\mathrm{c}}-\omega_{\mathrm{a}%
}>0$ two peaks are seen in the cavity transmission, the upper one corresponds
to the case where the qubit mainly occupies the ground state, whereas the
lower one, which is weaker, corresponds to the case where the qubit mainly
occupies the first excited state. The lower peak is less visible in the data
seen in panel a, which was obtained at lower input power, due to reduced
signal to noise ratio. In panels c and d the primary (labeled by $\pm1)$ and
superharmonic (labeled by $\pm2$, $\pm3$ and $4$) resonances are indicated by
arrows.}%
\label{Fig VNA}%
\end{center}
\end{figure*}

The chip is enclosed inside a copper package, which is cooled by a dilution
refrigerator to a temperature of $T=23%
\operatorname{mK}%
$. Both passive and active shielding methods are employed to suppress magnetic field
noise. While passive shielding is performed using a three-layer high
permeability metal, an active magnetic field compensation system placed
outside the cryostat is used to actively reduce low-frequency magnetic field
noise. A set of superconducting coils is used to apply DC magnetic flux. Qubit
state control is performed using shaped microwave pulses. Attenuators and filters
are installed at different cooling stages along the transmission lines for
qubit control and readout. A detailed description of sample fabrication and
experimental setup can be found in \cite{Orgiazzi_1407_1346,Bal_1324}.

The main theoretical results needed for analyzing the experimental findings
are briefly described below [derivations are given in the supplemental
material (SM)]. The circulating current states of the qubit are labeled as
$\left\vert \curvearrowleft\right\rangle $ and $\left\vert \curvearrowright
\right\rangle $. The coupling between the cavity mode and the qubit is
described by the term $-g\left(  A+A^{\dag}\right)  \left(  \left\vert
\curvearrowleft\right\rangle \left\langle \curvearrowleft\right\vert
-\left\vert \curvearrowright\right\rangle \left\langle \curvearrowright
\right\vert \right)  $ in the system Hamiltonian, where $A$ ($A^{\dag}$) is a cavity
mode annihilation (creation) operator, and $g$ is the coupling
coefficient. In the presence of an externally applied magnetic flux, the
energy gap $\hbar\omega_{\mathrm{a}}$ between the qubit ground state
$\left\vert -\right\rangle $ and first excited state $\left\vert
+\right\rangle $ is taken to be given by $\hbar\omega_{\mathrm{a}}=\hbar\sqrt
{\omega_{\mathrm{f}}^{2}+\omega_{\Delta}^{2}}$ (see SM section I.A), where%
\begin{equation}
\omega_{\mathrm{f}}=\frac{2I_{\mathrm{cc}}\Phi_{0}}{\hbar}\left(  \frac
{\Phi_{\mathrm{e}}}{\Phi_{0}}-\frac{1}{2}\right)  \;,
\end{equation}
$I_{\mathrm{cc}}$ ($-I_{\mathrm{cc}}$) is the circulating current associated
with the state $\left\vert \curvearrowright\right\rangle $ ($\left\vert
\curvearrowleft\right\rangle $), $\Phi_{0}=h/2e$ is the flux quantum,
$\Phi_{\mathrm{e}}$ is the externally applied magnetic flux and $\hbar
\omega_{\Delta}$ is the qubit energy gap for the case where $\Phi_{\mathrm{e}%
}/\Phi_{0}=1/2$.

The decoupled cavity mode is characterized by an angular resonance frequency
$\omega_{\mathrm{c}}$, Kerr coefficient $K_{\mathrm{c}}$, linear damping rate
$\gamma_{\mathrm{c}}$ and cubic damping (two-photon absorption) rate
$\gamma_{\mathrm{c}4}$. The response of the decoupled cavity in the weak
nonlinear regime (in which, nonlinearity is taken into account to lowest non-vanishing order) can be described by introducing the complex and mode amplitude
dependent cavity angular resonance frequency $\Upsilon_{\mathrm{c}}$, which is
given by%
\begin{equation}
\Upsilon_{\mathrm{c}}=\omega_{\mathrm{c}}-i\gamma_{\mathrm{c}}+\left(
K_{\mathrm{c}}-i\gamma_{\mathrm{c}4}\right)  E_{\mathrm{c}}\;,
\end{equation}
where $E_{\mathrm{c}}$ is the averaged number of photons occupying the cavity
mode. The imaginary part of $\Upsilon_{\mathrm{c}}$ represents the effect of
damping and the terms proportional to $E_{\mathrm{c}}$ determine the weak
nonlinear response. The effect of the flux qubit on the cavity response in the
weak nonlinear regime is theoretically evaluated in sections I and II of the SM
for the case where $g/\left\vert \omega_{\mathrm{c}}-\omega_{\mathrm{a}%
}\right\vert \ll1$. The coupling between the cavity mode and the qubit gives
rise to a resonance splitting. The steady state cavity mode response for the
case where the qubit mainly occupies the state $\left\vert \pm\right\rangle $
(ground and first excited states) is found to be equivalent to the response of
a mode having effective complex cavity angular resonance frequency
$\Upsilon_{\mathrm{eff}}$ given by
\begin{equation}
\Upsilon_{\mathrm{eff}}=\Upsilon_{\mathrm{c}}\pm\omega_{\mathrm{BS}}%
\pm\Upsilon_{\mathrm{ba}}\;, \label{Upsilon_eff}%
\end{equation}
where $\omega_{\mathrm{BS}}=g_{1}^{2}/\left(  \omega_{\mathrm{c}}%
+\omega_{\mathrm{a}}\right)  $ is the Bloch-Siegert shift \cite{Forn_237001}
(see SM section II). The term $\Upsilon_{\mathrm{ba}}$ is given by (see SM
section I.I)%
\begin{equation}
\Upsilon_{\mathrm{ba}}=-\frac{g_{1}^{2}}{\Delta_{1}}\frac{1-\frac{i}%
{\Delta_{1}T_{2}}}{1+\frac{1}{\Delta_{1}^{2}T_{2}^{2}}+\frac{4g_{1}^{2}%
T_{1}E_{\mathrm{c}}}{\Delta_{1}^{2}T_{2}}}\;, \label{Upsilon_ba bp}%
\end{equation}
$g_{1}=g/\beta_{\mathrm{f}}$ is the flux dependent effective coupling
coefficient, where the coefficient $\beta_{\mathrm{f}}$ is given by%
\begin{equation}
\beta_{\mathrm{f}}=\sqrt{1+\left(  \frac{\omega_{\mathrm{f}}}{\omega_{\Delta}%
}\right)  ^{2}}\;,
\end{equation}
$\Delta_{1}=\omega_{\mathrm{p}}-\omega_{\mathrm{a}}$ is the detuning between
the angular frequency of the externally injected pump tone $\omega
_{\mathrm{p}}$ and the qubit angular resonance frequency $\omega_{\mathrm{a}}%
$, and $T_{1}$ ($T_{2}$) is the qubit longitudinal (transverse) relaxation
time. Note that when $\Delta_{1}T_{2}\gg1$ and when the qubit mainly occupies
the state $\left\vert \pm\right\rangle $ the term $\Upsilon_{\mathrm{ba}}$
gives rise to a shift in the mode angular frequency approximately given by
$\mp g_{1}^{2}/\Delta_{1}$ and a shift in the value of the Kerr coefficient
approximately given by $\pm\left(  g_{1}^{4}/\Delta_{1}^{3}\right)  \left(
4T_{1}/T_{2}\right)  $. Similar theoretical results have been obtained in Ref. \cite{Boissonneault_060305}, in which the unitary
transformation that diagonalizes the Hamiltonian of the closed system has been
applied to the system's master equation.

The effect of the qubit on cavity response is experimentally investigated
using transmission measurements. The color coded plots in Fig. \ref{Fig VNA}
exhibit the measured (panels a and c) and calculated (panels b and d) cavity
transmission (in dB units) vs. $\omega_{\mathrm{f}}/2\pi$, for the case where
the power injected into the cavity is $-127$ dBm (panels a and b) and $-112$
dBm (panels c and d). In the first step of the theoretical calculation, which
has generated the theoretical predictions plotted in panels b and d, fixed
points are found by calculating steady state solutions of the equations of
motion that govern the dynamics of the system (see SM section I.I). Then in the
second step input-output relations are employed in order to calculate the
cavity transmission (see SM section I.E) \cite{Gardiner_3761}. The assumed
device parameters are listed in the caption of Fig. \ref{Fig VNA}.

While the cavity response seen in panels a and b of Fig. \ref{Fig VNA} is nearly linear,
nonlinearity is observed in the results depicted in panels c and d, which are
obtained at higher input power. The measured response exhibits hardening
(softening) when $\Delta_{1}<0$ ($\Delta_{1}>0$), for the case where the qubit
mainly occupies its ground state. The opposite behavior is obtained when the
qubit mainly occupies the first excited state. The probability for this to happen, which depends on the ratio between thermal energy and qubit energy gap, is non-negligible in the current experiment.  The comparison between the
experimental results (panels a and c) and the theoretical predictions (panels b
and d, respectively) yields an acceptable agreement.

It is well known that the flux qubit is expected to strongly affect the
response of the cavity mode near the primary resonance, i.e. when the ratio
$\omega_{\mathrm{a}}/\omega_{\mathrm{c}}$ is tuned close to unity (see the
points labeled by $\pm1$ in panels c and d of Fig. \ref{Fig VNA}). With
sufficiently large driving amplitude, however, higher order nonlinear
processes may give rise to superharmonic resonances, which occur near the
points at which the ratio $\omega_{\mathrm{a}}/\omega_{\mathrm{c}}$ is an
integer larger than unity (see the points labeled by $\pm2$, $\pm3$ and $4$ in
panels c and d of Fig. \ref{Fig VNA}). The cavity response near a
superharmonic resonance is theoretically evaluated in section III of the SM. It
is found that the same Eq. (\ref{Upsilon_ba bp}) can be used to describe the
effect of the flux qubit on cavity response near a superharmonic resonance,
provided that the coupling coefficient $g_{1}$ is replaced by $g_{n}$, which
is given by (see SM section III)%
\begin{equation}
g_{n}=g_{1}J_{1-n}\left(  \frac{4g_{1}\omega_{\mathrm{f}}E_{\mathrm{c}}^{1/2}%
}{\omega_{\mathrm{p}}\omega_{\Delta}}\right)  \;,
\end{equation}
where $J_{l}$ is the $l$'th Bessel function of the first kind, and the
detuning $\Delta_{1}=\omega_{\mathrm{p}}-\omega_{\mathrm{a}}$ is replaced by
$\Delta_{n}=n\omega_{\mathrm{p}}-\omega_{\mathrm{a}}$, where at the
superharmonic resonance $\omega_{\mathrm{a}}/\omega_{\mathrm{c}}=n$. As can be
seen by comparing panels c and d of Fig. \ref{Fig VNA}, the calculated and
measured cavity response near the superharmonic resonances exhibit an
acceptable agreement.%

\begin{figure}
[ptb]
\begin{center}
\includegraphics[
height=3.2397in,
width=3.4399in
]%
{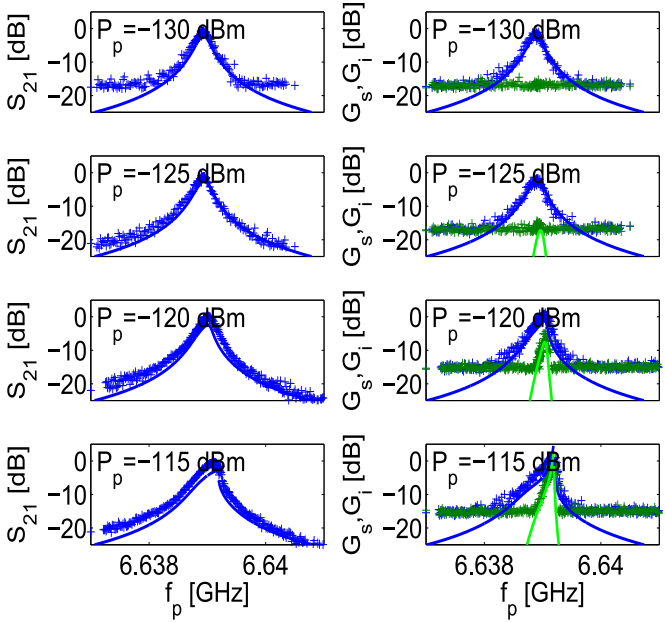}%
\caption{(color online) Cavity transmission (panels on the left) and IMD gain (panels on the
right). Experimental data points are labeled by crosses whereas the solid lines
represent the theoretical predictions (see SM section I.L) for the cavity
transmission $S_{21}$, for the signal gain $G_{\mathrm{s}}$ (blue) and for the
idler gain $G_{\mathrm{i}}$ (green). The parameters that have been employed
for the calculation are listed in the caption of Fig. \ref{Fig VNA}. The
detuning between the signal and pump frequencies is $\left(  \omega
_{\mathrm{s}}-\omega_{\mathrm{p}}\right)  /2\pi=50\operatorname{kHz}$.}%
\label{Fig IMD}%
\end{center}
\end{figure}

In general, nonlinear cavity response is commonly employed for frequency
mixing, which in turn can be used for signal amplification
\cite{Castellanos_083509,Vijay_110502,Castellanos_929,Bergeal_64,Hatridge_134501}
and noise squeezing \cite{Castellanos_929,Yurke_5054}. An amplifier based on flux qubits has been recently demonstrated in Ref. \cite{Rehak_162604}. Here we employ the
method of IMD to characterize frequency mixing. In this method, two
monochromatic tones are combined and injected into the cavity: an intense pump
tone at angular frequency $\omega_{\mathrm{p}}$ and amplitude $b_{\mathrm{c}%
1}^{\mathrm{in}}$, and a weaker signal tone at angular frequency
$\omega_{\mathrm{s}}=\omega_{\mathrm{p}}+\omega$ and amplitude $c_{\mathrm{c}%
1}^{\mathrm{in}}$. The cavity transmission is measured and the spectral
amplitude of the output signal tone at frequency $\omega_{\mathrm{s}}$, which
is labeled by $c_{\mathrm{c}2}^{\mathrm{out}}\left(  \omega\right)  $, and the
spectral amplitude of the so-called idler tone at frequency $2\omega
_{\mathrm{p}}-\omega_{\mathrm{s}}=\omega_{\mathrm{p}}-\omega$, which is
labeled by $c_{\mathrm{c}2}^{\mathrm{out}}\left(  -\omega\right)  $, are
recorded. The corresponding signal gain $G_{\mathrm{s}}=\left\vert
c_{\mathrm{c}2}^{\mathrm{out}}\left(  \omega\right)  /c_{\mathrm{c}%
1}^{\mathrm{in}}\right\vert ^{2}$ and idler gain $G_{\mathrm{i}}=\left\vert
c_{\mathrm{c}2}^{\mathrm{out}}\left(  -\omega\right)  /c_{\mathrm{c}%
1}^{\mathrm{in}}\right\vert ^{2}$ are determined, and the experimental
findings are compared with the theoretical predictions, which are based on the
linearized equations of motion of the system (see SM section I.L).

The results are exhibited in Fig. \ref{Fig IMD}, in which the cavity
transmission $S_{21}$ (left panels) and the signal $G_{\mathrm{s}}$ and idler
$G_{\mathrm{i}}$ gains (right panels) are plotted vs. pump frequency
$f_{\mathrm{p}}=\omega_{\mathrm{p}}/2\pi$ for different values of the pump
input power $P_{\mathrm{p}}$. The magnetic flux for these measurements is set
to a value for which $\omega_{\mathrm{f}}/2\pi=8.1%
\operatorname{GHz}%
$ and $\Delta_{1}/2\pi=-1.5%
\operatorname{GHz}%
$. Relatively good agreement between data and theory (see SM section I.L) is
found for the results seen in Fig. \ref{Fig IMD}, however, the deviation
between data and theory becomes larger at higher powers. Further study is
needed in order to identify the sources of discrepancy, and to improve the
accuracy of the theoretical predictions accordingly.

In summary, superharmonic resonances in the device under study have been
experimentally observed. We theoretically show that a relatively simple CQED
model of a system composed of two coupled elements, a single cavity mode
having no intrinsic nonlinearity and a two-level system, can account for the
main experimental findings. Further study will aim at expanding the range of
validity of the theoretical predictions in order to account for the
experimental results at higher levels of input power. Future experiments will
explore the possibility of exploiting nonlinearity for improving the fidelity
of qubit readout and employ superharmonic resonances for generating highly
correlated states of the microwave cavity field (e.g. the creation of
entangled pairs of microwave photons near the \thinspace$n=2$ superharmonic
resonance via two-photon stimulated emission events).

We thank Feyruz Kitapli and Pol Forn-D{\'\i}az for useful discussions. This research is supported by
the Gerald Schwartz and Heather Reisman Foundation, NSERC, Industry Canada,
CMC, CFI, Ontario MRI and the Israeli Science Foundation.

\bibliographystyle{apsrev}
\bibliography{C:/Users/eyal/software/swp40/TCITeX/BibTeX/bib/Eyal_bib/Eyal_Bib}

\end{document}


\title{Superharmonic Resonances in a Strongly Coupled Cavity-Atom System -
Supplemental Material}
\author{Eyal Buks}
\affiliation{Department of Electrical Engineering, Technion, Haifa 32000 Israel}
\author{Chunqing Deng}
\affiliation{Institute for Quantum Computing, University of Waterloo, Waterloo, ON, Canada N2L 3G1}
\affiliation{Department of Physics and Astronomy, University of Waterloo, Waterloo, ON, Canada N2L 3G1}
\affiliation{Waterloo Institute for Nanotechnology, University of Waterloo, Waterloo, ON, Canada N2L 3G1}
\author{Jean-Luc F.X. Orgazzi}
\affiliation{Institute for Quantum Computing, University of Waterloo, Waterloo, ON, Canada N2L 3G1}
\affiliation{Department of Electrical and Computer Engineering, University of Waterloo, Waterloo, ON, Canada N2L 3G1}
\affiliation{Waterloo Institute for Nanotechnology, University of Waterloo, Waterloo, ON, Canada N2L 3G1}
\author{Martin Otto}
\affiliation{Institute for Quantum Computing, University of Waterloo, Waterloo, ON, Canada N2L 3G1}
\affiliation{Department of Physics and Astronomy, University of Waterloo, Waterloo, ON, Canada N2L 3G1}
\affiliation{Waterloo Institute for Nanotechnology, University of Waterloo, Waterloo, ON, Canada N2L 3G1}
\author{Adrian Lupascu}
\affiliation{Institute for Quantum Computing, University of Waterloo, Waterloo, ON, Canada N2L 3G1}
\affiliation{Department of Physics and Astronomy, University of Waterloo, Waterloo, ON, Canada N2L 3G1}
\affiliation{Waterloo Institute for Nanotechnology, University of Waterloo, Waterloo, ON, Canada N2L 3G1}
\date{\today }
\maketitle





In the first section of the supplemental material the equations of motion are
derived and linearized and the response is evaluated in the weak nonlinear
regime. The second section is devoted to the Bloch-Siegert shift, and the
third one discusses the superharmonic resonances.

\section{Weak Nonlinear Response}

\subsection{The Closed System}

The Hamiltonian $\mathcal{H}_{0}$ of the closed system, formed by the flux
qubit and the cavity mode, is taken to be given by%
\begin{align}
\hbar^{-1}\mathcal{H}_{0}  &  =\omega_{\mathrm{c}}\left(  A^{\dag}A+\frac
{1}{2}\right)  +\frac{K_{\mathrm{c}}}{2}A^{\dagger}A^{\dagger}AA\nonumber\\
&  +\frac{\omega_{\mathrm{f}}}{2}\left(  \left\vert \curvearrowleft
\right\rangle \left\langle \curvearrowleft\right\vert -\left\vert
\curvearrowright\right\rangle \left\langle \curvearrowright\right\vert \right)
\nonumber\\
&  +\frac{\omega_{\Delta}}{2}\left(  \left\vert \curvearrowleft\right\rangle
\left\langle \curvearrowright\right\vert +\left\vert \curvearrowright
\right\rangle \left\langle \curvearrowleft\right\vert \right) \nonumber\\
&  -g\left(  A+A^{\dag}\right)  \left(  \left\vert \curvearrowleft
\right\rangle \left\langle \curvearrowleft\right\vert -\left\vert
\curvearrowright\right\rangle \left\langle \curvearrowright\right\vert
\right)  \ .\nonumber\\
&
\end{align}
The cavity mode angular resonance frequency is labeled by $\omega_{\mathrm{c}%
}$, $K_{\mathrm{c}}$ is the cavity mode Kerr coefficient and $A$ is the cavity
mode annihilation operator. The coefficient $\hbar\omega_{\mathrm{f}}$ is
related to the externally applied magnetic flux $\Phi_{\mathrm{e}}$ by%
\begin{equation}
\frac{\hbar\omega_{\mathrm{f}}}{2}=\frac{I_{\mathrm{cc}}\Phi_{0}}{2\pi}%
\phi_{\mathrm{e}}\;,
\end{equation}
where $I_{\mathrm{cc}}=-\left\langle \curvearrowright\right\vert
\partial\mathcal{H}_{0}/\partial\Phi_{\mathrm{e}}\left\vert \curvearrowright
\right\rangle $ ($-I_{\mathrm{cc}}$) is the circulating current associated
with the state $\left\vert \curvearrowright\right\rangle $ ($\left\vert
\curvearrowleft\right\rangle $), $\Phi_{0}=h/2e$ is the flux quantum, and the
normalized applied magnetic flux $\phi_{\mathrm{e}}$ is given by%
\begin{equation}
\phi_{\mathrm{e}}=2\pi\left(  \frac{\Phi_{\mathrm{e}}}{\Phi_{0}}-\frac{1}%
{2}\right)  \;.
\end{equation}
The coefficient $\hbar\omega_{\Delta}$ is the qubit energy gap, and $g$ is the
coupling constant. The frequencies $\omega_{\mathrm{f}}$, $\omega_{\Delta}$
and $g$ are assumed to be time independent.

\subsection{Qubit Energy Eigenstates}

The energy eigenstates of the decoupled qubit $\left\vert \pm\right\rangle $
are given by%
\begin{equation}
\left(
\begin{array}
[c]{c}%
\left\vert +\right\rangle \\
\left\vert -\right\rangle
\end{array}
\right)  =\left(
\begin{array}
[c]{cc}%
\cos\frac{\theta}{2} & \sin\frac{\theta}{2}\\
-\sin\frac{\theta}{2} & \cos\frac{\theta}{2}%
\end{array}
\right)  \left(
\begin{array}
[c]{c}%
\left\vert \curvearrowleft\right\rangle \\
\left\vert \curvearrowright\right\rangle
\end{array}
\right)  \ ,
\end{equation}
where%
\begin{equation}
\tan\theta=\frac{\omega_{\Delta}}{\omega_{\mathrm{f}}}\ , \label{tan theta}%
\end{equation}
and the corresponding eigenenergies are%
\begin{equation}
\varepsilon_{\pm}=\pm\frac{\hbar\omega_{\mathrm{a}}}{2}\ ,
\end{equation}
where%
\begin{equation}
\omega_{\mathrm{a}}=\sqrt{\omega_{\mathrm{f}}^{2}+\omega_{\Delta}^{2}}\ .
\label{omega_a}%
\end{equation}
The following relations%
\begin{equation}
\left\vert \curvearrowleft\right\rangle \left\langle \curvearrowleft
\right\vert -\left\vert \curvearrowright\right\rangle \left\langle
\curvearrowright\right\vert =\cos\theta\;\Sigma_{z}-\sin\theta\left(
\Sigma_{+}+\Sigma_{-}\right)  \ ,
\end{equation}
and%
\begin{equation}
\left\vert \curvearrowleft\right\rangle \left\langle \curvearrowright
\right\vert +\left\vert \curvearrowright\right\rangle \left\langle
\curvearrowleft\right\vert =\sin\theta\;\Sigma_{z}+\cos\theta\left(
\Sigma_{+}+\Sigma_{-}\right)  \ ,
\end{equation}
 hold, where%
\begin{align}
\Sigma_{z}  &  =\left\vert +\right\rangle \left\langle +\right\vert
-\left\vert -\right\rangle \left\langle -\right\vert \;,\\
\Sigma_{+}  &  =\left\vert +\right\rangle \left\langle -\right\vert \;,\\
\Sigma_{-}  &  =\left\vert -\right\rangle \left\langle +\right\vert \;,
\end{align}
and thus the Hamiltonian $\mathcal{H}_{0}$ can be expressed as%
\begin{align}
\hbar^{-1}\mathcal{H}_{0}  &  =\omega_{\mathrm{c}}\left(  A^{\dag}A+\frac
{1}{2}\right)  +\frac{K_{\mathrm{c}}}{2}A^{\dagger}A^{\dagger}AA\nonumber\\
&  +\frac{\omega_{\mathrm{a}}}{2}\Sigma_{z}\nonumber\\
&  -g\left(  A+A^{\dag}\right)  \left[  \cos\theta\;\Sigma_{z}-\sin
\theta\left(  \Sigma_{+}+\Sigma_{-}\right)  \right]  \ .\nonumber\\
&  \label{H_0 Sigma}%
\end{align}

\subsection{Damping}

Damping is taken into account using a model containing reservoirs having dense
spectrum of oscillator modes interacting with both the cavity mode and the
qubit. The cavity mode is assumed to be coupled to 4 semi-infinite
transmission lines. The first two, denoted as $\mathrm{c}1$ and $\mathrm{c}2$,
are feedlines (or ports), which are linearly coupled to the cavity mode with
coupling magnitudes $\gamma_{\mathrm{c}1}$ and $\gamma_{\mathrm{c}2}$ and
coupling phases $\phi_{\mathrm{c}1}$ and $\phi_{\mathrm{c}2}$, respectively,
and which are employed to deliver the input and output signals. The third,
denoted as $\mathrm{c}3$, is linearly coupled to the cavity mode with a
coupling magnitude $\gamma_{\mathrm{c}3}$ and a coupling phase $\phi
_{\mathrm{c}3}$, and it is used to model linear dissipation (due to internal
sources), whereas the forth one, denoted as $\mathrm{c}4$, is nonlinearly
coupled to the cavity mode with a coupling magnitude $\gamma_{\mathrm{c}4}$
and a coupling phase $\phi_{\mathrm{c}4}$, and is employed to model nonlinear
dissipation (due to internal sources). The qubit is assumed to be coupled to 2
semi-infinite transmission lines, with coupling magnitudes $\gamma
_{\mathrm{q}1}$ and $\gamma_{\mathrm{q}2}$ and coupling phases $\phi
_{\mathrm{q}1}$ and $\phi_{\mathrm{q}2}$, respectively. While the first is
employed to model energy relaxation, the second is employed to model
dephasing. Note that all coupling parameters are assumed to be frequency
independent. The following Bose%
\begin{align}
\left[  A,A^{\dagger}\right]   &  =1\;,\\
\left[  a_{\mathrm{c}n}\left(  \omega\right)  ,a_{\mathrm{c}m}^{\dagger
}\left(  \omega^{\prime}\right)  \right]   &  =\delta_{n,m}\delta\left(
\omega-\omega^{\prime}\right)  \;,\\
\left[  a_{\mathrm{q}n}\left(  \omega\right)  ,a_{\mathrm{q}m}^{\dagger
}\left(  \omega^{\prime}\right)  \right]   &  =\delta_{n,m}\delta\left(
\omega-\omega^{\prime}\right)  \;,\\
\left[  a_{\mathrm{c}n}\left(  \omega\right)  ,a_{\mathrm{c}m}\left(
\omega^{\prime}\right)  \right]   &  =0\;,\\
\left[  a_{\mathrm{q}n}\left(  \omega\right)  ,a_{\mathrm{q}m}\left(
\omega^{\prime}\right)  \right]   &  =0\;,
\end{align}
and qubit%
\begin{align}
\left[  \Sigma_{z},\Sigma_{+}\right]   &  =2\Sigma_{+}\;,\\
\left[  \Sigma_{z},\Sigma_{-}\right]   &  =-2\Sigma_{-}\;,\\
\left[  \Sigma_{+},\Sigma_{-}\right]   &  =\Sigma_{z}\;,
\end{align}
commutation relations are assumed to hold.

The Hamiltonian $\mathcal{H}$ of the system is taken to be given by%
\begin{align}
\hbar^{-1}\mathcal{H}  &  =\hbar^{-1}\mathcal{H}_{0}\nonumber\\
&  +\sum_{n=1}^{4}\int\mathrm{d}\omega\;\omega a_{\mathrm{c}n}^{\dagger
}\left(  \omega\right)  a_{\mathrm{c}n}\left(  \omega\right) \nonumber\\
&  +\sum_{n=1}^{3}\sqrt{\frac{\gamma_{\mathrm{c}n}}{\pi}}\int\mathrm{d}%
\omega\;\left[  e^{i\phi_{\mathrm{c}n}}A^{\dagger}a_{\mathrm{c}n}\left(
\omega\right)  +e^{-i\phi_{\mathrm{c}n}}a_{\mathrm{c}n}^{\dagger}\left(
\omega\right)  A\right] \nonumber\\
&  +\sqrt{\frac{\gamma_{\mathrm{c}4}}{2\pi}}\int\mathrm{d}\omega\;\left[
e^{i\phi_{\mathrm{c}4}}A^{\dagger}A^{\dagger}a_{\mathrm{c}4}\left(
\omega\right)  +e^{-i\phi_{\mathrm{c}4}}a_{\mathrm{c}4}^{\dagger}\left(
\omega\right)  AA\right] \nonumber\\
&  +\sum_{n=1}^{2}\int\mathrm{d}\omega\;\omega a_{\mathrm{q}n}^{\dagger
}\left(  \omega\right)  a_{\mathrm{q}n}\left(  \omega\right) \nonumber\\
&  +\sqrt{\frac{\gamma_{\mathrm{q}1}}{2\pi}}\int\mathrm{d}\omega\left(
e^{i\phi_{\mathrm{q}1}}\Sigma_{+}a_{\mathrm{q}1}\left(  \omega\right)
+e^{-i\phi_{\mathrm{q}1}}a_{\mathrm{q}1}^{\dagger}\left(  \omega\right)
\Sigma_{-}\right) \nonumber\\
&  +\sqrt{\frac{\gamma_{\mathrm{q}2}}{4\pi}}\int\mathrm{d}\omega\left(
e^{i\phi_{\mathrm{q}2}}\Sigma_{z}a_{\mathrm{q}2}\left(  \omega\right)
+e^{-i\phi_{\mathrm{q}2}}a_{\mathrm{q}2}^{\dagger}\left(  \omega\right)
\Sigma_{z}\right)  \;.\nonumber\\
&
\end{align}

\subsection{The equations of motion}

The Heisenberg equations of motion are generated according to
\begin{equation}
\frac{\mathrm{d}O}{\mathrm{d}t}=-i\left[  O,\hbar^{-1}\mathcal{H}\right]  \;,
\end{equation}
where $O$ is an operator and $\mathcal{H}$ is the total Hamiltonian, hence%
\begin{align}
\frac{\mathrm{d}A}{\mathrm{d}t}  &  =-i\omega_{\mathrm{c}}A-iK_{\mathrm{c}%
}A^{\dagger}AA\nonumber\\
&  +ig\left[  \cos\theta\;\Sigma_{z}-\sin\theta\left(  \Sigma_{+}+\Sigma
_{-}\right)  \right] \nonumber\\
&  -i\sum_{n=1}^{3}\sqrt{\frac{\gamma_{\mathrm{c}n}}{\pi}}e^{i\phi
_{\mathrm{c}n}}\int\mathrm{d}\omega\;a_{\mathrm{c}n}\left(  \omega\right)
\nonumber\\
&  -i\sqrt{\frac{2\gamma_{\mathrm{c}4}}{\pi}}e^{i\phi_{\mathrm{c}4}}%
\int\mathrm{d}\omega\;A^{\dagger}a_{\mathrm{c}4}\left(  \omega\right)
\;,\nonumber\\
&  \label{dA/dt V1}%
\end{align}%
\begin{align}
\frac{\mathrm{d}\Sigma_{z}}{\mathrm{d}t}  &  =-2ig\sin\theta\left(  A+A^{\dag
}\right)  \left(  \Sigma_{+}-\Sigma_{-}\right) \nonumber\\
&  -2i\sqrt{\frac{\gamma_{\mathrm{q}1}}{2\pi}}\int\mathrm{d}\omega\nonumber\\
&  \times\left(  e^{i\phi_{\mathrm{q}1}}\Sigma_{+}a_{\mathrm{q}1}\left(
\omega\right)  -e^{-i\phi_{\mathrm{q}1}}a_{\mathrm{q}1}^{\dagger}\left(
\omega\right)  \Sigma_{-}\right)  \;,\nonumber\\
&  \label{d Sigma_z/dt V1}%
\end{align}%
\begin{align}
\frac{\mathrm{d}\Sigma_{+}}{\mathrm{d}t}  &  =i\omega_{\mathrm{a}}\Sigma
_{+}-ig\left(  A+A^{\dag}\right)  \left(  2\cos\theta\;\Sigma_{+}+\sin
\theta\;\Sigma_{z}\right) \nonumber\\
&  -i\sqrt{\frac{\gamma_{\mathrm{q}1}}{2\pi}}\int\mathrm{d}\omega
e^{-i\phi_{\mathrm{q}1}}a_{\mathrm{q}1}^{\dagger}\left(  \omega\right)
\Sigma_{z}\nonumber\\
&  +i\sqrt{\frac{\gamma_{\mathrm{q}2}}{\pi}}\int\mathrm{d}\omega\nonumber\\
&  \times\left(  e^{i\phi_{\mathrm{q}2}}\Sigma_{+}a_{\mathrm{q}2}\left(
\omega\right)  +e^{-i\phi_{\mathrm{q}2}}a_{\mathrm{q}2}^{\dagger}\left(
\omega\right)  \Sigma_{+}\right)  \;,\nonumber\\
&  \label{d Sigma_+/dt V1}%
\end{align}%
\begin{align}
&  \frac{\mathrm{d}a_{\mathrm{c}n}\left(  \omega\right)  }{\mathrm{d}%
t}\nonumber\\
&  =\left\{
\begin{array}
[c]{cc}%
-i\omega a_{\mathrm{c}n}\left(  \omega\right)  -i\sqrt{\frac{\gamma
_{\mathrm{c}n}}{\pi}}e^{-i\phi_{\mathrm{c}n}}A & n=1,2,3\\
-i\omega a_{\mathrm{c}4}\left(  \omega\right)  -i\sqrt{\frac{\gamma
_{\mathrm{c}4}}{2\pi}}e^{-i\phi_{\mathrm{c}4}}AA & n=4
\end{array}
\right.  \ ,\nonumber\\
&  \label{da_cn/dt}%
\end{align}%
\begin{equation}
\frac{\mathrm{d}a_{\mathrm{q}1}\left(  \omega\right)  }{\mathrm{d}t}=-i\omega
a_{\mathrm{q}1}\left(  \omega\right)  -i\sqrt{\frac{\gamma_{\mathrm{q}1}}%
{2\pi}}e^{-i\phi_{\mathrm{q}1}}\Sigma_{-}\ , \label{da_q1/dt}%
\end{equation}
and%
\begin{equation}
\frac{\mathrm{d}a_{\mathrm{q}2}\left(  \omega\right)  }{\mathrm{d}t}=-i\omega
a_{\mathrm{q}2}\left(  \omega\right)  -i\sqrt{\frac{\gamma_{\mathrm{q}2}}%
{4\pi}}e^{-i\phi_{\mathrm{q}2}}\Sigma_{z}\;. \label{da_q2/dt}%
\end{equation}

\subsection{Input-Output Relations}

The field operator $a_{\mathrm{c}n}\left(  t,\omega\right)  $ at time $t$ can
be calculated by assuming either initial value for the field operator
$a_{\mathrm{c}n}\left(  t_{0},\omega\right)  $ at time $t_{0}$ or final value
for the field operator $a_{\mathrm{c}n}\left(  t_{1},\omega\right)  $ at time
$t_{1}$. The time $t_{0}$ is assumed to be in the distant past whereas $t_{1}$
is assumed to be in the distant future, i.e. $t_{0}\ll t\ll t_{1}$. Time
integration of (\ref{da_cn/dt}) using initial values at time $t_{0}<t$ yields%
\begin{align}
&  a_{\mathrm{c}n}\left(  \omega\right) \nonumber\\
&  =\left\{
\begin{array}
[c]{cc}%
\begin{array}
[c]{c}%
e^{-i\omega\left(  t-t_{0}\right)  }a_{\mathrm{c}n}\left(  t_{0},\omega\right)
\\
-i\sqrt{\frac{\gamma_{\mathrm{c}n}}{\pi}}e^{-i\phi_{\mathrm{c}n}}\int_{t_{0}%
}^{t}\mathrm{d}t^{\prime}\;e^{-i\omega\left(  t-t^{\prime}\right)  }A\left(
t^{\prime}\right)
\end{array}
& n=1,2,3\\%
\begin{array}
[c]{c}%
e^{-i\omega\left(  t-t_{0}\right)  }a_{\mathrm{c}4}\left(  t_{0},\omega\right)
\\
-i\sqrt{\frac{\gamma_{\mathrm{c}4}}{2\pi}}e^{-i\phi_{\mathrm{c}4}}\int_{t_{0}%
}^{t}\mathrm{d}t^{\prime}\;e^{-i\omega\left(  t-t^{\prime}\right)  }A\left(
t^{\prime}\right)  A\left(  t^{\prime}\right)
\end{array}
& n=4
\end{array}
\right.  \ ,\nonumber\\
&
\end{align}
and using finite values at time $t_{1}>t$ yields%
\begin{align}
&  a_{\mathrm{c}n}\left(  \omega\right) \nonumber\\
&  =\left\{
\begin{array}
[c]{cc}%
\begin{array}
[c]{c}%
e^{-i\omega\left(  t-t_{1}\right)  }a_{\mathrm{c}n}\left(  t_{1},\omega\right)
\\
-i\sqrt{\frac{\gamma_{\mathrm{c}n}}{\pi}}e^{-i\phi_{\mathrm{c}n}}\int_{t_{1}%
}^{t}\mathrm{d}t^{\prime}\;e^{-i\omega\left(  t-t^{\prime}\right)  }A\left(
t^{\prime}\right)
\end{array}
& n=1,2,3\\%
\begin{array}
[c]{c}%
e^{-i\omega\left(  t-t_{1}\right)  }a_{\mathrm{c}4}\left(  t_{1},\omega\right)
\\
-i\sqrt{\frac{\gamma_{\mathrm{c}4}}{2\pi}}e^{-i\phi_{\mathrm{c}4}}\int_{t_{1}%
}^{t}\mathrm{d}t^{\prime}\;e^{-i\omega\left(  t-t^{\prime}\right)  }A\left(
t^{\prime}\right)  A\left(  t^{\prime}\right)
\end{array}
& n=4
\end{array}
\right.  \ .\nonumber\\
&
\end{align}
Integrating $a_{\mathrm{c}n}\left(  \omega\right)  $ over $\omega$ and using
the following relations
\begin{equation}
\int_{-\infty}^{\infty}\mathrm{d}\omega\;e^{-i\omega\left(  t-t^{\prime
}\right)  }=2\pi\delta\left(  t-t^{\prime}\right)  \;,
\end{equation}
and%
\begin{equation}
\int_{t_{\mathrm{c}}}^{t}\mathrm{d}t^{\prime}\;\delta\left(  t-t^{\prime
}\right)  f\left(  t^{\prime}\right)  =\frac{1}{2}\mathrm{sgn}\left(
t-t_{\mathrm{c}}\right)  f\left(  t\right)  \ ,
\end{equation}
where $\mathrm{sgn}\left(  x\right)  $ is the sign function
\begin{equation}
\mathrm{sgn}\left(  x\right)  =\left\{
\begin{array}
[c]{cc}%
+1 & \mathrm{if}\ x>0\\
-1 & \mathrm{if}\ x<0.
\end{array}
\right.  \ ,
\end{equation}
lead to%
\begin{align}
&  \frac{1}{\sqrt{2\pi}}\int_{-\infty}^{\infty}\mathrm{d}\omega\;a_{\mathrm{c}%
n}\left(  \omega\right) \nonumber\\
&  =\left\{
\begin{array}
[c]{cc}%
a_{\mathrm{c}n}^{\mathrm{in}}\left(  t\right)  -i\sqrt{\frac{\gamma
_{\mathrm{c}n}}{2}}e^{-i\phi_{\mathrm{c}n}}A\left(  t\right)  & n=1,2,3\\
a_{\mathrm{c}4}^{\mathrm{in}}\left(  t\right)  -i\frac{\sqrt{\gamma
_{\mathrm{c}4}}}{2}e^{-i\phi_{\mathrm{c}4}}A\left(  t\right)  A\left(
t\right)  & n=4
\end{array}
\right.  \ ,\nonumber\\
&  \label{a_n omega in}%
\end{align}
and%
\begin{align}
&  \frac{1}{\sqrt{2\pi}}\int_{-\infty}^{\infty}\mathrm{d}\omega\;a_{\mathrm{c}%
n}\left(  \omega\right) \nonumber\\
&  =\left\{
\begin{array}
[c]{cc}%
a_{\mathrm{c}n}^{\mathrm{out}}\left(  t\right)  +i\sqrt{\frac{\gamma
_{\mathrm{c}n}}{2}}e^{-i\phi_{\mathrm{c}n}}A\left(  t\right)  & n=1,2,3\\
a_{\mathrm{c}4}^{\mathrm{out}}\left(  t\right)  +i\frac{\sqrt{\gamma
_{\mathrm{c}4}}}{2}e^{-i\phi_{\mathrm{c}4}}A\left(  t\right)  A\left(
t\right)  & n=4
\end{array}
\right.  \ ,\nonumber\\
&  \label{a_n omega out}%
\end{align}
where the input operators are given by%
\begin{equation}
a_{\mathrm{c}n}^{\mathrm{in}}\left(  t\right)  =\frac{1}{\sqrt{2\pi}}%
\int_{-\infty}^{\infty}\mathrm{d}\omega\;e^{-i\omega\left(  t-t_{0}\right)
}a_{\mathrm{c}n}\left(  t_{0},\omega\right)  \ ,
\end{equation}
and the output operators by
\begin{equation}
a_{\mathrm{c}n}^{\mathrm{out}}\left(  t\right)  =\frac{1}{\sqrt{2\pi}}%
\int_{-\infty}^{\infty}\mathrm{d}\omega\;e^{-i\omega\left(  t-t_{1}\right)
}a_{\mathrm{c}n}\left(  t_{1},\omega\right)  \ .
\end{equation}
Equations (\ref{a_n omega in}) and (\ref{a_n omega out}) yield the following
input-output relations%
\begin{align}
&  a_{\mathrm{c}n}^{\mathrm{out}}\left(  t\right)  -a_{\mathrm{c}%
n}^{\mathrm{in}}\left(  t\right) \nonumber\\
&  =\left\{
\begin{array}
[c]{cc}%
-i\sqrt{2\gamma_{\mathrm{c}n}}e^{-i\phi_{\mathrm{c}n}}A\left(  t\right)  &
n=1,2,3\\
-i\sqrt{\gamma_{\mathrm{c}4}}e^{-i\phi_{\mathrm{c}4}}A\left(  t\right)
A\left(  t\right)  & n=4
\end{array}
\right.  \ .\nonumber\\
&  \label{out-in}%
\end{align}

Similarly for the bath operators that are coupled to the qubit one has [see
Eqs. (\ref{da_q1/dt}) and (\ref{da_q2/dt})]%
\begin{equation}
\frac{1}{\sqrt{2\pi}}\int_{-\infty}^{\infty}\mathrm{d}\omega\;a_{\mathrm{q}%
1}\left(  \omega\right)  =a_{\mathrm{q}1}^{\mathrm{in}}\left(  t\right)
-i\sqrt{\frac{\gamma_{\mathrm{q}1}}{4}}e^{-i\phi_{\mathrm{q}1}}\Sigma_{-}\ ,
\label{int a_q1}%
\end{equation}
where%
\begin{equation}
a_{\mathrm{q}1}^{\mathrm{in}}\left(  t\right)  =\frac{1}{\sqrt{2\pi}}%
\int_{-\infty}^{\infty}\mathrm{d}\omega\;e^{-i\omega\left(  t-t_{0}\right)
}a_{\mathrm{q}1}\left(  t_{0},\omega\right)  \ ,
\end{equation}
and%
\begin{equation}
\frac{1}{\sqrt{2\pi}}\int_{-\infty}^{\infty}\mathrm{d}\omega\;a_{\mathrm{q}%
2}\left(  \omega\right)  =a_{\mathrm{q}2}^{\mathrm{in}}\left(  t\right)
-i\sqrt{\frac{\gamma_{\mathrm{q}2}}{8}}e^{-i\phi_{\mathrm{q}2}}\Sigma_{z}\ ,
\label{int a_q2}%
\end{equation}
where%
\begin{equation}
a_{\mathrm{q}2}^{\mathrm{in}}\left(  t\right)  =\frac{1}{\sqrt{2\pi}}%
\int_{-\infty}^{\infty}\mathrm{d}\omega\;e^{-i\omega\left(  t-t_{0}\right)
}a_{\mathrm{q}2}\left(  t_{0},\omega\right)  \ .
\end{equation}

Thus, the equation of motion for $A$ becomes [see Eqs. (\ref{dA/dt V1}) and
(\ref{a_n omega in})]%
\begin{align}
&  \frac{\mathrm{d}A}{\mathrm{d}t}+\left[  i\omega_{\mathrm{c}}+\gamma
_{\mathrm{c}}+\left(  iK_{\mathrm{c}}+\gamma_{\mathrm{c}4}\right)  A^{\dagger
}A\right]  A\nonumber\\
&  =ig\left[  \cos\theta\;\Sigma_{z}-\sin\theta\left(  \Sigma_{+}+\Sigma
_{-}\right)  \right] \nonumber\\
&  -i\sum_{n=1}^{3}\sqrt{2\gamma_{\mathrm{c}n}}e^{i\phi_{\mathrm{c}n}%
}a_{\mathrm{c}n}^{\mathrm{in}}-2i\sqrt{\gamma_{\mathrm{c}4}}e^{i\phi
_{\mathrm{c}4}}A^{\dag}a_{\mathrm{c}4}^{\mathrm{in}}\;,\nonumber\\
&  \label{dA/dt V2}%
\end{align}
where%
\begin{equation}
\gamma_{\mathrm{c}}=\gamma_{\mathrm{c}1}+\gamma_{\mathrm{c}2}+\gamma
_{\mathrm{c}3}\;.
\end{equation}
Furthermore, by making use of the following relations%
\begin{align}
\Sigma_{+}\Sigma_{-}  &  =\frac{1}{2}\left(  1+\Sigma_{z}\right)  \;,\\
\Sigma_{-}\Sigma_{+}  &  =\frac{1}{2}\left(  1-\Sigma_{z}\right)  \;,\\
\Sigma_{z}\Sigma_{+}  &  =-\Sigma_{+}\Sigma_{z}=\Sigma_{+}\;,\\
\Sigma_{-}\Sigma_{z}  &  =-\Sigma_{z}\Sigma_{-}=\Sigma_{-}\;,
\end{align}
one finds that the equation of motion for $\Sigma_{z}$ becomes [see Eqs.
(\ref{d Sigma_z/dt V1}) and (\ref{int a_q1})]%
\begin{align}
&  \frac{\mathrm{d}\Sigma_{z}}{\mathrm{d}t}+\gamma_{\mathrm{q}1}\left(
1+\Sigma_{z}\right)  +2ig\sin\theta\left(  A+A^{\dag}\right)  \left(
\Sigma_{+}-\Sigma_{-}\right) \nonumber\\
&  =2i\sqrt{\gamma_{\mathrm{q}1}}\left(  -\Sigma_{+}e^{i\phi_{\mathrm{q}1}%
}a_{\mathrm{q}1}^{\mathrm{in}}+e^{-i\phi_{\mathrm{q}1}}a_{\mathrm{q}%
1}^{\mathrm{in}\dag}\Sigma_{-}\right)  \;,\nonumber\\
&  \label{d Sigma_z/dt V2}%
\end{align}
and the equation of motion for $\Sigma_{+}$ becomes [see Eqs.
(\ref{d Sigma_+/dt V1}) and (\ref{int a_q2})]%
\begin{align}
&  \frac{\mathrm{d}\Sigma_{+}}{\mathrm{d}t}-i\omega_{\mathrm{a}}\Sigma
_{+}+\left(  \frac{\gamma_{\mathrm{q}1}}{2}+\gamma_{\mathrm{q}2}\right)
\Sigma_{+}\nonumber\\
&  +ig\left(  A+A^{\dag}\right)  \left(  2\cos\theta\;\Sigma_{+}+\sin
\theta\;\Sigma_{z}\right) \nonumber\\
&  =-i\sqrt{\gamma_{\mathrm{q}1}}e^{-i\phi_{\mathrm{q}1}}a_{\mathrm{q}%
1}^{\mathrm{in}\dag}\Sigma_{z}\nonumber\\
&  +i\sqrt{2\gamma_{\mathrm{q}2}}\left(  \Sigma_{+}e^{i\phi_{\mathrm{q}2}%
}a_{\mathrm{q}2}^{\mathrm{in}}+e^{-i\phi_{\mathrm{q}2}}a_{\mathrm{q}%
2}^{\mathrm{in}\dag}\Sigma_{+}\right)  \;.\nonumber\\
&  \label{d Sigma_+/dt V2}%
\end{align}

\subsection{Cavity External Drive}

Consider the case where a monochromatic pump tone having amplitude
$b_{\mathrm{c}1}^{\mathrm{in}}$ and angular frequency $\omega_{\mathrm{p}}$ is
injected into port $1$. In a frame rotating at angular frequency
$\omega_{\mathrm{p}}$ the input cavity operators are expressed as%
\begin{equation}
a_{\mathrm{c}n}^{\mathrm{in}}=\left\{
\begin{array}
[c]{cc}%
\left(  b_{\mathrm{c}n}^{\mathrm{in}}+c_{\mathrm{c}n}^{\mathrm{in}}\right)
e^{-i\omega_{\mathrm{p}}t} & n=1\\
c_{\mathrm{c}n}^{\mathrm{in}}e^{-i\omega_{\mathrm{p}}t} & n=2,3,4
\end{array}
\right.  \ ,
\end{equation}
the input qubit operators as%
\begin{equation}
a_{\mathrm{q}n}^{\mathrm{in}}=c_{\mathrm{q}n}^{\mathrm{in}}e^{-i\omega
_{\mathrm{p}}t}\ ,
\end{equation}
the output cavity operators as%
\begin{equation}
a_{\mathrm{c}n}^{\mathrm{out}}=\left(  b_{\mathrm{c}n}^{\mathrm{out}%
}+c_{\mathrm{c}n}^{\mathrm{out}}\right)  e^{-i\omega_{\mathrm{p}}t}\;,
\end{equation}
the cavity mode annihilation operator as%
\begin{equation}
A=A_{\mathrm{R}}e^{-i\omega_{\mathrm{p}}t}\;,
\end{equation}
and the qubit operator $\Sigma_{+}$ as%
\begin{equation}
\Sigma_{+}=\Sigma_{+\mathrm{R}}e^{i\omega_{\mathrm{p}}t}\;.
\end{equation}

In terms of these notations Eq. (\ref{dA/dt V2}) becomes%
\begin{align}
&  \frac{\mathrm{d}A_{\mathrm{R}}}{\mathrm{d}t}+\left[  -i\Delta_{\mathrm{pc}%
}+\gamma_{\mathrm{c}}+\left(  iK_{\mathrm{c}}+\gamma_{\mathrm{c}4}\right)
A_{\mathrm{R}}^{\dag}A_{\mathrm{R}}\right]  A_{\mathrm{R}}\nonumber\\
&  +i\sqrt{2\gamma_{\mathrm{c}1}}e^{i\phi_{\mathrm{c}1}}b_{\mathrm{c}%
1}^{\mathrm{in}}\nonumber\\
&  -ig\left[  \cos\theta\;\Sigma_{z}e^{i\omega_{\mathrm{p}}t}-\sin
\theta\left(  \Sigma_{+\mathrm{R}}e^{2i\omega_{\mathrm{p}}t}+\Sigma
_{+\mathrm{R}}^{\dag}\right)  \right] \nonumber\\
&  =\mathcal{V}_{\mathrm{A}}\;,\nonumber\\
&  \label{dA/dt R}%
\end{align}
where%
\begin{equation}
\Delta_{\mathrm{pc}}=\omega_{\mathrm{p}}-\omega_{\mathrm{c}}\;,
\end{equation}
and where%
\begin{equation}
\mathcal{V}_{\mathrm{A}}=-i\sum_{n=1}^{3}\sqrt{2\gamma_{\mathrm{c}n}}%
e^{i\phi_{\mathrm{c}n}}c_{\mathrm{c}n}^{\mathrm{in}}-2i\sqrt{\gamma
_{\mathrm{c}4}}e^{i\left(  \phi_{\mathrm{c}4}+\omega_{\mathrm{p}}t\right)
}A_{\mathrm{R}}^{\dag}c_{\mathrm{c}4}^{\mathrm{in}}\;, \label{V_A}%
\end{equation}
Eq. (\ref{d Sigma_z/dt V2}) becomes%
\begin{align}
&  \frac{\mathrm{d}\Sigma_{z}}{\mathrm{d}t}+\gamma_{\mathrm{q}1}\left(
1+\Sigma_{z}\right) \nonumber\\
&  +2ig\sin\theta\left(  A_{\mathrm{R}}e^{-i\omega_{\mathrm{p}}t}%
+A_{\mathrm{R}}^{\dag}e^{i\omega_{\mathrm{p}}t}\right)  \left(  \Sigma
_{+\mathrm{R}}e^{i\omega_{\mathrm{p}}t}-\Sigma_{+\mathrm{R}}^{\dag}%
e^{-i\omega_{\mathrm{p}}t}\right) \nonumber\\
&  =\mathcal{V}_{\mathrm{z}}\;,\nonumber\\
&  \label{d Sigma_z/dt R}%
\end{align}
where%
\begin{equation}
\mathcal{V}_{\mathrm{z}}=2i\sqrt{\gamma_{\mathrm{q}1}}\left(  -e^{i\phi
_{\mathrm{q}1}}\Sigma_{+\mathrm{R}}c_{\mathrm{q}1}^{\mathrm{in}}%
+e^{-i\phi_{\mathrm{q}1}}c_{\mathrm{q}1}^{\mathrm{in}\dag}\Sigma_{+\mathrm{R}%
}^{\dag}\right)  \;, \label{V_z}%
\end{equation}
and Eq. (\ref{d Sigma_+/dt V2}) becomes%
\begin{align}
&  \frac{\mathrm{d}\Sigma_{+\mathrm{R}}}{\mathrm{d}t}+i\Delta_{1}%
\Sigma_{+\mathrm{R}}+\left(  \frac{\gamma_{\mathrm{q}1}}{2}+\gamma
_{\mathrm{q}2}\right)  \Sigma_{+\mathrm{R}}\nonumber\\
&  +ig\left(  A_{\mathrm{R}}e^{-i\omega_{\mathrm{p}}t}+A_{\mathrm{R}}^{\dag
}e^{i\omega_{\mathrm{p}}t}\right)  \left(  2\cos\theta\;\Sigma_{+\mathrm{R}%
}+\sin\theta\;\Sigma_{z}e^{-i\omega_{\mathrm{p}}t}\right) \nonumber\\
&  =\mathcal{V}_{\mathrm{+}}\;,\nonumber\\
&  \label{d Sigma_+=/dt R}%
\end{align}
where%
\begin{equation}
\Delta_{1}=\omega_{\mathrm{p}}-\omega_{\mathrm{a}}\;,
\end{equation}
and where%
\begin{align}
\mathcal{V}_{\mathrm{+}}  &  =-i\sqrt{\gamma_{\mathrm{q}1}}e^{-i\phi
_{\mathrm{q}1}}c_{\mathrm{q}1}^{\mathrm{in}\dag}\Sigma_{z}\nonumber\\
&  +i\sqrt{2\gamma_{\mathrm{q}2}}\left(  e^{i\phi_{\mathrm{q}2}}%
\Sigma_{+\mathrm{R}}c_{\mathrm{q}2}^{\mathrm{in}}e^{-i\omega_{\mathrm{p}}%
t}+e^{-i\phi_{\mathrm{q}2}}c_{\mathrm{q}2}^{\mathrm{in}\dag}\Sigma
_{+\mathrm{R}}e^{i\omega_{\mathrm{p}}t}\right)  \;.\nonumber\\
&  \label{V_+}%
\end{align}

\subsection{Rotating Wave Approximation}

In the rotating wave approximation (RWA), in which rapidly oscillating terms
are disregarded, the equations of motion (\ref{dA/dt R}),
(\ref{d Sigma_z/dt R}) and (\ref{d Sigma_+=/dt R}) become%
\begin{align}
&  \frac{\mathrm{d}A_{\mathrm{R}}}{\mathrm{d}t}+\left[  -i\Delta_{\mathrm{pc}%
}+\gamma_{\mathrm{c}}+\left(  iK_{\mathrm{c}}+\gamma_{\mathrm{c}4}\right)
A_{\mathrm{R}}^{\dag}A_{\mathrm{R}}\right]  A_{\mathrm{R}}\nonumber\\
&  +i\sqrt{2\gamma_{\mathrm{c}1}}e^{i\phi_{\mathrm{c}1}}b_{\mathrm{c}%
1}^{\mathrm{in}}+ig_{1}\Sigma_{+\mathrm{R}}^{\dag}=\mathcal{V}_{\mathrm{A}%
}\;,\nonumber\\
&  \label{dA_R/dt RWA}%
\end{align}%
\begin{align}
&  \frac{\mathrm{d}\Sigma_{z}}{\mathrm{d}t}+\gamma_{\mathrm{q}1}\left(
1+\Sigma_{z}\right) \nonumber\\
+2ig_{1}\left(  A_{\mathrm{R}}\Sigma_{+\mathrm{R}}-\Sigma_{+\mathrm{R}}^{\dag
}A_{\mathrm{R}}^{\dag}\right)   &  =\mathcal{V}_{\mathrm{z}}\;,\nonumber\\
&  \label{d Sigma_z/dt RWA}%
\end{align}
and%
\begin{align}
&  \frac{\mathrm{d}\Sigma_{+\mathrm{R}}}{\mathrm{d}t}+i\Delta_{1}%
\Sigma_{+\mathrm{R}}+\left(  \frac{\gamma_{\mathrm{q}1}}{2}+\gamma
_{\mathrm{q}2}\right)  \Sigma_{+\mathrm{R}}\nonumber\\
+ig_{1}A_{\mathrm{R}}^{\dag}\Sigma_{z}  &  =\mathcal{V}_{\mathrm{+}%
}\;,\nonumber\\
&  \label{d Sigma_+/dt RWA}%
\end{align}
where%
\begin{equation}
g_{1}=g\sin\theta\;. \label{g=g_0*sin(theta)}%
\end{equation}

\subsection{Linearization}

Expectation values of the operators $\mathcal{V}_{\mathrm{A}}$, $\mathcal{V}%
_{\mathrm{z}}$ and $\mathcal{V}_{\mathrm{+}}$ are evaluated by assuming that
bath modes are all in thermal equilibrium \cite{Gardiner_3761}. To first order
in the damping coefficients one finds that $\left\langle \mathcal{V}%
_{\mathrm{A}}\right\rangle $ vanishes [see Eq. (\ref{V_A})] and that [see Eqs.
(\ref{V_z}) and (\ref{V_+})]%
\begin{equation}
\left\langle \mathcal{V}_{\mathrm{z}}\right\rangle =-2\gamma_{\mathrm{q}%
1}n_{0}\left\langle \Sigma_{z}\right\rangle \;, \label{<V_z>=}%
\end{equation}%
\begin{equation}
\left\langle \mathcal{V}_{\mathrm{+}}\right\rangle =-2\left(  \frac
{\gamma_{\mathrm{q}1}}{2}+\gamma_{\mathrm{q}2}\right)  n_{0}\left\langle
\Sigma_{+\mathrm{R}}\right\rangle \;, \label{<V_+>=}%
\end{equation}
where $n_{0}$ is the Bosonic thermal occupation number.

With the help of Eqs. (\ref{dA_R/dt RWA}), (\ref{d Sigma_z/dt RWA}),
(\ref{d Sigma_+/dt RWA}), (\ref{<V_z>=}) and (\ref{<V_+>=}) the equations of
motion become%
\begin{equation}
\frac{\mathrm{d}A_{\mathrm{R}}}{\mathrm{d}t}+\Theta_{\mathrm{R}}%
=\mathcal{F}_{\mathrm{A}}\;,
\end{equation}%
\begin{equation}
\frac{\mathrm{d}\Sigma_{z}}{\mathrm{d}t}+\Theta_{\mathrm{z}}=\mathcal{F}%
_{\mathrm{z}}\;,
\end{equation}
and%
\begin{equation}
\frac{\mathrm{d}\Sigma_{+\mathrm{R}}}{\mathrm{d}t}+\Theta_{+}=\mathcal{F}%
_{\mathrm{+}}\;,
\end{equation}
where%
\begin{align}
&  \Theta_{\mathrm{R}}\left(  A_{\mathrm{R}},A_{\mathrm{R}}^{\dag},\Sigma
_{z},\Sigma_{+\mathrm{R}},\Sigma_{+\mathrm{R}}^{\dag}\right) \nonumber\\
&  =\left[  -i\Delta_{\mathrm{pc}}+\gamma_{\mathrm{c}}+\left(  iK_{\mathrm{c}%
}+\gamma_{\mathrm{c}4}\right)  A_{\mathrm{R}}^{\dag}A_{\mathrm{R}}\right]
A_{\mathrm{R}}\nonumber\\
&  +i\sqrt{2\gamma_{\mathrm{c}1}}e^{i\phi_{\mathrm{c}1}}b_{\mathrm{c}%
1}^{\mathrm{in}}+ig_{1}\Sigma_{+\mathrm{R}}^{\dag}\;,\nonumber\\
&  \label{Theta_R}%
\end{align}%
\begin{align}
&  \Theta_{\mathrm{z}}\left(  A_{\mathrm{R}},A_{\mathrm{R}}^{\dag},\Sigma
_{z},\Sigma_{+\mathrm{R}},\Sigma_{+\mathrm{R}}^{\dag}\right) \nonumber\\
&  =\frac{\Sigma_{z}-P_{0}}{T_{1}}+2ig_{1}\left(  A_{\mathrm{R}}%
\Sigma_{+\mathrm{R}}-\Sigma_{+\mathrm{R}}^{\dag}A_{\mathrm{R}}^{\dag}\right)
\;,\nonumber\\
&  \label{Theta_z}%
\end{align}
and%
\begin{align}
&  \Theta_{+}\left(  A_{\mathrm{R}},A_{\mathrm{R}}^{\dag},\Sigma_{z}%
,\Sigma_{+\mathrm{R}},\Sigma_{+\mathrm{R}}^{\dag}\right) \nonumber\\
&  =\frac{\Sigma_{+\mathrm{R}}}{T_{2}}+i\Delta_{1}\Sigma_{+\mathrm{R}}%
+ig_{1}A_{\mathrm{R}}^{\dag}\Sigma_{z}\;.\nonumber\\
&  \label{Theta_+}%
\end{align}
The forcing terms $\mathcal{F}_{\mathrm{A}}=\mathcal{V}_{\mathrm{A}%
}-\left\langle \mathcal{V}_{\mathrm{A}}\right\rangle $, $\mathcal{F}%
_{\mathrm{z}}=\mathcal{V}_{\mathrm{z}}-\left\langle \mathcal{V}_{\mathrm{z}%
}\right\rangle $ and $\mathcal{F}_{+}=\mathcal{V}_{+}-\left\langle
\mathcal{V}_{+}\right\rangle $ have a vanishing thermal expectation value. The
coefficient $P_{0}$, which is given by%
\begin{equation}
P_{0}=-\frac{1}{2n_{0}+1}\;,
\end{equation}
represents the expectation value $\left\langle \Sigma_{z}\right\rangle $ in
thermal equilibrium in the absent of external driving and when the coupling
between the qubit and the cavity can be disregarded. The time $T_{1}$, which
is given by%
\begin{equation}
T_{1}=-\frac{P_{0}}{\gamma_{\mathrm{q}1}}\;,
\end{equation}
is the qubit longitudinal relaxation time, and the time $T_{2}$, which is
given by%
\begin{equation}
T_{2}=-\frac{P_{0}}{\frac{\gamma_{\mathrm{q}1}}{2}+\gamma_{\mathrm{q}2}}%
=\frac{2T_{1}}{1+\frac{2\gamma_{\mathrm{q}2}}{\gamma_{\mathrm{q}1}}}\;,
\end{equation}
is the qubit transverse relaxation time.

\subsection{Fixed Points}

The solution is expressed as%
\begin{subequations}
\begin{align}
A_{\mathrm{R}}  &  =\alpha_{\mathrm{R}}+a_{\mathrm{R}}\;,\\
\Sigma_{z}  &  =P_{z}+\sigma_{z}\;,\\
\Sigma_{+\mathrm{R}}  &  =P_{+\mathrm{R}}+\sigma_{+\mathrm{R}}\;,
\end{align}
where both $\alpha_{\mathrm{R}}$ and $P_{+\mathrm{R}}$ are complex numbers,
$P_{z}$ is a real number, and the operators $a_{\mathrm{R}}$, $\sigma_{z}$ and
$\sigma_{+\mathrm{R}}$ are considered as small. Fixed points are found by
solving%
\end{subequations}
\begin{subequations}
\begin{align}
\Theta_{\mathrm{R}}\left(  \alpha_{\mathrm{R}},\alpha_{\mathrm{R}}^{\ast
},P_{z},P_{+\mathrm{R}},P_{+\mathrm{R}}^{\ast}\right)   &
=0\;,\label{Theta_R=0}\\
\Theta_{\mathrm{z}}\left(  \alpha_{\mathrm{R}},\alpha_{\mathrm{R}}^{\ast
},P_{z},P_{+\mathrm{R}},P_{+\mathrm{R}}^{\ast}\right)   &
=0\;,\label{Theta_z=0}\\
\Theta_{+}\left(  \alpha_{\mathrm{R}},\alpha_{\mathrm{R}}^{\ast}%
,P_{z},P_{+\mathrm{R}},P_{+\mathrm{R}}^{\ast}\right)   &  =0\;.
\label{Theta_+=0}%
\end{align}
The solution of $\Theta_{\mathrm{z}}=\Theta_{+}=0$ yields%
\end{subequations}
\begin{equation}
P_{+\mathrm{R}}=-\frac{ig_{1}T_{2}\alpha_{\mathrm{R}}^{\ast}P_{z}}%
{1+i\Delta_{1}T_{2}}\;,
\end{equation}
and%
\begin{equation}
P_{0}=\left(  1+\frac{4g_{1}^{2}T_{1}T_{2}\left\vert \alpha_{\mathrm{R}%
}\right\vert ^{2}}{1+\Delta_{1}^{2}T_{2}^{2}}\right)  P_{z}\;, \label{P_0}%
\end{equation}
and thus%
\begin{equation}
P_{+\mathrm{R}}=-\frac{ig_{1}T_{2}\alpha_{\mathrm{R}}^{\ast}\left(
1-i\Delta_{1}T_{2}\right)  P_{0}}{1+\Delta_{1}^{2}T_{2}^{2}+4g_{1}^{2}%
T_{1}T_{2}E_{\mathrm{c}}}\;, \label{P_+R}%
\end{equation}
where%
\begin{equation}
E_{\mathrm{c}}=\left\vert \alpha_{\mathrm{R}}\right\vert ^{2}\;.
\end{equation}
Substituting into the condition $\Theta_{\mathrm{R}}=0$ yields%
\begin{align}
0  &  =\left[  -i\Delta_{\mathrm{pc}}+\gamma_{\mathrm{c}}+\left(
iK_{\mathrm{c}}+\gamma_{\mathrm{c}4}\right)  E_{\mathrm{c}}+i\Upsilon
_{\mathrm{ba}}P_{0}\right]  \alpha_{\mathrm{R}}\nonumber\\
&  +i\sqrt{2\gamma_{\mathrm{c}1}}e^{i\phi_{\mathrm{c}1}}b_{\mathrm{c}%
1}^{\mathrm{in}}\;,\nonumber\\
&  \label{eq for alpha_R}%
\end{align}
where%
\begin{equation}
\Upsilon_{\mathrm{ba}}=\frac{g_{1}^{2}T_{2}\left(  i-\Delta_{1}T_{2}\right)
}{1+\Delta_{1}^{2}T_{2}^{2}+4g_{1}^{2}T_{1}T_{2}E_{\mathrm{c}}}\;.
\label{Upsilon_ba}%
\end{equation}
or%
\begin{align}
\Upsilon_{\mathrm{ba}}  &  =-\frac{g_{1}^{2}}{\Delta_{1}}\frac{1-i\zeta_{2}%
}{1+\zeta_{2}^{2}+\frac{4g_{1}^{2}\zeta_{2}E_{\mathrm{c}}}{\Delta_{1}^{2}%
\zeta_{1}}}\nonumber\\
&  =-\frac{g_{1}^{2}}{\Delta_{1}}\frac{1-i\zeta_{2}}{1+\zeta_{2}^{2}%
}\nonumber\\
&  -\frac{4ig_{1}^{4}}{\Delta_{1}^{3}}\frac{\zeta_{2}\left(  i+\zeta
_{2}\right)  }{\zeta_{1}\left(  1+\zeta_{2}^{2}\right)  ^{2}}E_{\mathrm{c}%
}\nonumber\\
&  +O\left(  E_{\mathrm{c}}^{2}\right)  \;,\nonumber\\
&  \label{xi_ba}%
\end{align}
where%
\begin{equation}
\zeta_{n}=\frac{1}{\Delta_{1}T_{n}}\;,
\end{equation}
and where $n\in\left\{  1,2\right\}  $, thus to second order in $\left\vert
\alpha_{\mathrm{R}}\right\vert $ Eq. (\ref{eq for alpha_R}) can be expressed
as%
\begin{equation}
0=\left(  i\Omega+\Gamma\right)  \alpha_{\mathrm{R}}+i\sqrt{2\gamma
_{\mathrm{c}1}}e^{i\phi_{\mathrm{c}1}}b_{\mathrm{c}1}^{\mathrm{in}}\;,
\label{eq for alpha_R V1}%
\end{equation}
where%
\begin{align}
\Omega &  =\Omega_{0}+\Omega_{2}E_{\mathrm{c}}\;,\\
\Gamma &  =\Gamma_{0}+\Gamma_{2}E_{\mathrm{c}}\;,
\end{align}
and where%
\begin{align}
\Omega_{0}  &  =-\Delta_{\mathrm{pc}}-\frac{g_{1}^{2}}{\Delta_{1}}\frac{P_{0}%
}{1+\zeta_{2}^{2}}\;,\label{Omega_0}\\
\Omega_{2}  &  =K_{\mathrm{c}}+\frac{4g_{1}^{4}}{\Delta_{1}^{3}}\frac
{\zeta_{2}P_{0}}{\zeta_{1}\left(  1+\zeta_{2}^{2}\right)  ^{2}}%
\;,\label{Omega_2}\\
\Gamma_{0}  &  =\gamma_{\mathrm{c}}-\frac{g_{1}^{2}}{\Delta_{1}}\frac
{\zeta_{2}P_{0}}{1+\zeta_{2}^{2}}\;,\label{Gamma_0}\\
\Gamma_{2}  &  =\gamma_{\mathrm{c}4}+\frac{4g_{1}^{4}}{\Delta_{1}^{3}}%
\frac{\zeta_{2}^{2}P_{0}}{\zeta_{1}\left(  1+\zeta_{2}^{2}\right)  ^{2}}\;.
\label{Gamma_2}%
\end{align}

Taking the module squared of Eq. (\ref{eq for alpha_R V1}) leads to%
\begin{equation}
\left[  \left(  \Omega_{0}+\Omega_{2}E_{\mathrm{c}}\right)  ^{2}+\left(
\Gamma_{0}+\Gamma_{2}E_{\mathrm{c}}\right)  ^{2}\right]  E_{\mathrm{c}%
}=S_{\mathrm{p}}\;, \label{eq for E}%
\end{equation}
where%
\begin{equation}
S_{\mathrm{p}}=2\gamma_{\mathrm{c}1}\left\vert b_{\mathrm{c}1}^{\mathrm{in}%
}\right\vert ^{2}\;.
\end{equation}
Finding $E_{\mathrm{c}}$ by solving Eq. (\ref{eq for E}) allows calculating
$\alpha_{\mathrm{R}}$ according to Eq. (\ref{eq for alpha_R V1}), calculating
$P_{z}$ according to Eq. (\ref{P_0}) and calculating $P_{+\mathrm{R}}$
according to Eq. (\ref{P_+R}).

\subsection{Onset of Bistability Point}

In general, for any fixed value of the driving amplitude $S_{\mathrm{p}}$ Eq.
(\ref{eq for E}) can be expressed as a relation between $E_{\mathrm{c}}$ and
$\omega_{\mathrm{p}}$. When $S_{\mathrm{p}}$ is sufficiently large the
response of the system becomes bistable, that is $E_{\mathrm{c}}$ becomes a
multi-valued function of $\omega_{\mathrm{p}}$ in some range near resonance.
The onset of bistability point is defined as the point for which%
\begin{align}
\frac{\partial\Omega_{0}}{\partial E_{\mathrm{c}}}  &  =0\ ,\\
\frac{\partial^{2}\Omega_{0}}{\partial E_{\mathrm{c}}^{2}}  &  =0\ .
\end{align}
By solving the above conditions one finds that the values of $E_{\mathrm{c}}$,
$\Omega_{0}$ and $S_{\mathrm{p}}$ at the onset of bistability point, which are
labeled as $E_{\mathrm{c},\mathrm{o}}$, $\Omega_{0,\mathrm{o}}$ and
$S_{\mathrm{p},\mathrm{o}}$, respectively, are given by \cite{Yurke_5054}%
\begin{equation}
E_{\mathrm{c},\mathrm{o}}=\frac{2\Gamma_{0}}{\sqrt{3}\left(  \left\vert
\Omega_{2}\right\vert -\sqrt{3}\Gamma_{2}\right)  }\ , \label{E_c,0}%
\end{equation}%
\begin{equation}
\Omega_{0,\mathrm{o}}=-\Gamma_{0}\frac{\Omega_{2}}{\left\vert \Omega
_{2}\right\vert }\frac{4\Gamma_{2}|\Omega_{2}|+\sqrt{3}\left(  \Omega_{2}%
^{2}+\Gamma_{2}^{2}\right)  }{\Omega_{2}^{2}-3\Gamma_{2}^{2}}\ ,
\label{Omega_0,o}%
\end{equation}
and%
\begin{equation}
S_{\mathrm{p},\mathrm{o}}=\frac{8}{3\sqrt{3}}\frac{\Gamma_{0}^{3}(\Omega
_{2}^{2}+\Gamma_{2}^{2})}{\left(  \left\vert \Omega_{2}\right\vert -\sqrt
{3}\Gamma_{2}\right)  ^{3}}\ .
\end{equation}
Bistability is possible only when nonlinear damping is sufficiently small%
\begin{equation}
\Gamma_{2}<\frac{\left\vert \Omega_{2}\right\vert }{\sqrt{3}}\;.
\label{Gamma_2<}%
\end{equation}

\subsection{Susceptibility}

The linearized equations of motion can be expressed in a matrix form as%
\begin{equation}
\frac{\mathrm{d}}{\mathrm{d}t}\left(
\begin{array}
[c]{c}%
a_{\mathrm{R}}\\
a_{\mathrm{R}}^{\dag}\\
\sigma_{z}\\
\sigma_{+\mathrm{R}}\\
\sigma_{+\mathrm{R}}^{\dag}%
\end{array}
\right)  +J\left(
\begin{array}
[c]{c}%
a_{\mathrm{R}}\\
a_{\mathrm{R}}^{\dag}\\
\sigma_{z}\\
\sigma_{+\mathrm{R}}\\
\sigma_{+\mathrm{R}}^{\dag}%
\end{array}
\right)  =\left(
\begin{array}
[c]{c}%
\mathcal{F}_{\mathrm{A}}\\
\mathcal{F}_{\mathrm{A}}^{\dag}\\
\mathcal{F}_{\mathrm{z}}\\
\mathcal{F}_{\mathrm{+}}\\
\mathcal{F}_{\mathrm{+}}^{\dag}%
\end{array}
\right)  \;, \label{d/dt fluct}%
\end{equation}
where%
\begin{equation}
J=\frac{\partial\left(  \Theta_{\mathrm{R}},\Theta_{\mathrm{R}}^{\dag}%
,\Theta_{\mathrm{z}},\Theta_{+},\Theta_{+}^{\dag}\right)  }{\partial\left(
A_{\mathrm{R}},A_{\mathrm{R}}^{\dag},\Sigma_{z},\Sigma_{+\mathrm{R}}%
,\Sigma_{+\mathrm{R}}^{\dag}\right)  }\;
\end{equation}
is the Jacobian matrix [see Eqs. (\ref{Theta_R}), (\ref{Theta_z}) and
(\ref{Theta_+})], which is evaluated at a fixed point $\left(  \alpha
_{\mathrm{R}},\alpha_{\mathrm{R}}^{\ast},P_{z},P_{+\mathrm{R}},P_{+\mathrm{R}%
}^{\ast}\right)  $. The Jacobian matrix can be expressed as%
\begin{equation}
J=J_{0}+g_{1}V\;, \label{J=J_0+gV}%
\end{equation}
where $J_{0}$ is given in a block form by%
\begin{equation}
J_{0}=\left(
\begin{tabular}
[c]{c|c}%
$J_{0\mathrm{c}}$ & $0$\\\hline
$0$ & $J_{0\mathrm{q}}$%
\end{tabular}
\ \ \ \right)  \;,
\end{equation}
the $2\times2$ matrix $J_{0\mathrm{c}}$ is given by%
\begin{equation}
J_{0\mathrm{c}}=\left(
\begin{array}
[c]{cc}%
W & V\\
V^{\ast} & W^{\ast}%
\end{array}
\right)  \;,
\end{equation}
and the coefficients $W$ and $V$ are given by%
\begin{align}
W  &  =\frac{\partial\Theta_{\mathrm{R}}}{\partial A_{\mathrm{R}}}%
=-i\Delta_{\mathrm{pc}}+\gamma_{\mathrm{c}}+2\left(  iK_{\mathrm{c}}%
+\gamma_{\mathrm{c}4}\right)  E_{\mathrm{c}}\;,\\
V  &  =\frac{\partial\Theta_{\mathrm{R}}}{\partial A_{\mathrm{R}}^{\dag}%
}=\left(  iK_{\mathrm{c}}+\gamma_{\mathrm{c}4}\right)  \alpha_{\mathrm{R}}%
^{2}\;.
\end{align}
The diagonal $3\times3$ matrix $J_{0\mathrm{q}}$ is given by%
\begin{equation}
J_{0\mathrm{q}}=\left(
\begin{array}
[c]{ccc}%
\frac{1}{T_{1}} & 0 & 0\\
0 & \frac{1}{T_{2}}+i\Delta_{1} & 0\\
0 & 0 & \frac{1}{T_{2}}-i\Delta_{1}%
\end{array}
\right)  \;,
\end{equation}
and the matrix $V$ is given by%
\begin{align}
V  &  =i\left(
\begin{array}
[c]{ccccc}%
0 & 0 & 0 & 0 & 1\\
0 & 0 & 0 & -1 & 0\\
2P_{+\mathrm{R}} & -2P_{+\mathrm{R}}^{\ast} & 0 & 2\alpha_{\mathrm{R}} &
-2\alpha_{\mathrm{R}}^{\ast}\\
0 & P_{z} & \alpha_{\mathrm{R}}^{\ast} & 0 & 0\\
-P_{z} & 0 & -\alpha_{\mathrm{R}} & 0 & 0
\end{array}
\right)  \;.\nonumber\\
&
\end{align}

In general, the Fourier transform of a time dependent variable or operator
$O\left(  t\right)  $ is denoted as $O\left(  \omega\right)  $%
\begin{equation}
O\left(  t\right)  =\frac{1}{\sqrt{2\pi}}\int_{-\infty}^{\infty}%
\mathrm{d}\omega\;O\left(  \omega\right)  e^{-i\omega t}\;.
\end{equation}
Applying the Fourier transform to Eq. (\ref{d/dt fluct}) yields%
\begin{equation}
\left(
\begin{array}
[c]{c}%
a_{\mathrm{R}}\left(  \omega\right) \\
a_{\mathrm{R}}^{\dag}\left(  -\omega\right) \\
\sigma_{z}\left(  \omega\right) \\
\sigma_{+\mathrm{R}}\left(  \omega\right) \\
\sigma_{+\mathrm{R}}^{\dag}\left(  -\omega\right)
\end{array}
\right)  =\chi\left(  \omega\right)  \left(
\begin{array}
[c]{c}%
\mathcal{F}_{\mathrm{A}}\left(  \omega\right) \\
\mathcal{F}_{\mathrm{A}}^{\dag}\left(  -\omega\right) \\
\mathcal{F}_{\mathrm{z}}\left(  \omega\right) \\
\mathcal{F}_{\mathrm{+}}\left(  \omega\right) \\
\mathcal{F}_{\mathrm{+}}^{\dag}\left(  -\omega\right)
\end{array}
\right)  \;,
\end{equation}
where the susceptibility $\chi\left(  \omega\right)  $ is given by%
\begin{equation}
\chi\left(  \omega\right)  =\left(  J-i\omega\right)  ^{-1}\;. \label{chi}%
\end{equation}

The matrix $\chi_{0}\left(  \omega\right)  =\left(  J_{0}-i\omega\right)
^{-1}$ can be expressed in a block form as%
\begin{equation}
\chi_{0}\left(  \omega\right)  =\left(
\begin{tabular}
[c]{c|c}%
$\chi_{\mathrm{c}}\left(  \omega\right)  $ & $0$\\\hline
$0$ & $\chi_{\mathrm{q}}\left(  \omega\right)  $%
\end{tabular}
\ \ \right)  \;,
\end{equation}
where the cavity block $\chi_{\mathrm{c}}\left(  \omega\right)  =\left(
J_{0\mathrm{c}}-i\omega\right)  ^{-1}$ is given by%
\begin{equation}
\chi_{\mathrm{c}}\left(  \omega\right)  =\frac{\left(
\begin{array}
[c]{cc}%
W^{\ast}-i\omega & -V\\
-V^{\ast} & W-i\omega
\end{array}
\right)  }{\left(  \lambda_{1}-i\omega\right)  \left(  \lambda_{2}%
-i\omega\right)  }\;, \label{chi_c}%
\end{equation}
$\lambda_{1}$ and $\lambda_{2}$, which are given by%
\begin{align}
\lambda_{1}+\lambda_{2}  &  =W+W^{\ast}\;,\\
\lambda_{1}\lambda_{2}  &  =\left\vert W\right\vert ^{2}-\left\vert
V\right\vert ^{2}\;,
\end{align}
are the eigenvalues of $J_{0\mathrm{c}}$, and where the qubit block
$\chi_{\mathrm{q}}\left(  \omega\right)  $ is given by%
\begin{equation}
\chi_{\mathrm{q}}\left(  \omega\right)  =\left(  J_{0\mathrm{q}}%
-i\omega\right)  ^{-1}\;. \label{chi_q}%
\end{equation}

\subsection{Intermodulation}

In this section the output field of feedline $2$ is evaluated for the case
where, in addition to the pump, a monochromatic input signal is injected into
feedline $1$. Its amplitude $c_{\mathrm{c}1}^{\mathrm{in}}\left(
\omega\right)  $, as well as the resultant cavity mode amplitude
$a_{\mathrm{R}}\left(  \omega\right)  $ and output feedline amplitudes
$c_{\mathrm{c}n}^{\mathrm{out}}\left(  \omega\right)  $ are considered as
complex numbers (rather than operators). The phase $\phi_{\mathrm{c}1}$ is
assumed to vanish. With the help of the input-output relations given by Eq.
(\ref{out-in}) one finds that the meanfield amplitude $b_{2}^{\mathrm{out}}$
of the output signal of feedline $2$ is given by%
\begin{equation}
b_{\mathrm{c}2}^{\mathrm{out}}=-i\sqrt{2\gamma_{\mathrm{c}2}}e^{-i\phi
_{\mathrm{c}2}}\alpha_{\mathrm{R}}\ ,\label{b_2^out}%
\end{equation}
and the fluctuation amplitude $c_{\mathrm{c}2}^{\mathrm{out}}\left(
\omega\right)  $ is given by%
\begin{equation}
c_{\mathrm{c}2}^{\mathrm{out}}\left(  \omega\right)  =-i\sqrt{2\gamma
_{\mathrm{c}2}}e^{-i\phi_{\mathrm{c}2}}a_{\mathrm{R}}\left(  \omega\right)
\ .\label{c_c2^out}%
\end{equation}
In terms of the cavity-cavity $2\times2$ block of the susceptibility matrix
$\chi\left(  \omega\right)  $, which is denoted as $\chi_{\mathrm{cc}}\left(
\omega\right)  $, the cavity amplitude $a_{\mathrm{R}}\left(  \omega\right)  $
can be expressed as%
\begin{equation}
\left(
\begin{array}
[c]{c}%
a_{\mathrm{R}}\left(  \omega\right)  \\
a_{\mathrm{R}}^{\ast}\left(  -\omega\right)
\end{array}
\right)  =\sqrt{2\gamma_{\mathrm{c}1}}\chi_{\mathrm{cc}}\left(  \omega\right)
\left(
\begin{array}
[c]{c}%
-ic_{\mathrm{c}1}^{\mathrm{in}}\left(  \omega\right)  \\
ic_{\mathrm{c}1}^{\mathrm{in}\ast}\left(  -\omega\right)
\end{array}
\right)  \;,
\end{equation}
and thus [see Eq. (\ref{c_c2^out})]%
\begin{equation}
\left(
\begin{array}
[c]{c}%
c_{\mathrm{c}2}^{\mathrm{out}}\left(  \omega\right)  \\
c_{\mathrm{c}2}^{\mathrm{out}\dag}\left(  -\omega\right)
\end{array}
\right)  =\mathcal{R}_{\mathrm{cc}}\left(
\begin{array}
[c]{c}%
c_{\mathrm{c}1}^{\mathrm{in}}\left(  \omega\right)  \\
-c_{\mathrm{c}1}^{\mathrm{in}\ast}\left(  -\omega\right)
\end{array}
\right)  \;,
\end{equation}
where%
\begin{equation}
\mathcal{R}_{\mathrm{cc}}=2\sqrt{\gamma_{\mathrm{c}1}\gamma_{\mathrm{c}2}%
}\left(
\begin{array}
[c]{cc}%
-e^{-i\phi_{\mathrm{c}2}} & 0\\
0 & e^{i\phi_{\mathrm{c}2}}%
\end{array}
\right)  \chi_{\mathrm{cc}}\left(  \omega\right)  \;.
\end{equation}
The signal gain is defined by%
\begin{equation}
G_{\mathrm{s}}=\left\vert \frac{c_{\mathrm{c}2}^{\mathrm{out}}\left(
\omega\right)  }{c_{\mathrm{c}1}^{\mathrm{in}}\left(  \omega\right)
}\right\vert ^{2}\;,\label{G_s}%
\end{equation}
and the idler gain is defined by%
\begin{equation}
G_{\mathrm{i}}=\left\vert \frac{c_{\mathrm{c}2}^{\mathrm{out}}\left(
-\omega\right)  }{c_{\mathrm{c}1}^{\mathrm{in}}\left(  \omega\right)
}\right\vert ^{2}\;.\label{G_i}%
\end{equation}

\section{Bloch-Siegert Shift}

Consider the case where intrinsic cavity Kerr nonlinearity can be disregarded,
i.e. the case where $K_{\mathrm{c}}=0$. For that case the Hamiltonian of the
closed system $\mathcal{H}_{0}$ (\ref{H_0 Sigma}) can be expressed as%
\begin{equation}
\mathcal{H}_{0}=\mathcal{H}_{\mathrm{JC}}+\mathcal{V}_{\mathrm{BS}}\ ,
\end{equation}
where $\mathcal{H}_{\mathrm{JC}}$, which is given by%
\begin{align}
\hbar^{-1}\mathcal{H}_{\mathrm{JC}}  &  =\omega_{\mathrm{c}}\left(  A^{\dag
}A+\frac{1}{2}\right)  +\frac{\omega_{\mathrm{a}}}{2}\Sigma_{z}\nonumber\\
&  +g_{1}\left(  A^{\dag}\Sigma_{-}+A\Sigma_{+}\right)  \ ,\nonumber\\
&  \label{H JC}%
\end{align}
is the Jaynes-Cummings Hamiltonian, the term $\mathcal{V}_{\mathrm{BS}}$ is
given by%
\begin{equation}
\hbar^{-1}\mathcal{V}_{\mathrm{BS}}=g_{1}\left[  A\Sigma_{-}+\Sigma_{+}%
A^{\dag}-\left(  A+A^{\dag}\right)  \Sigma_{z}\cot\theta\right]  \ ,
\label{V_BS}%
\end{equation}
and $g_{1}$ is given by Eq. (\ref{g=g_0*sin(theta)}). In the rotating wave
approximation (RWA), in which rapidly oscillating terms are disregarded, the
term $\mathcal{V}_{\mathrm{BS}}$ is ignored.

The states $\left\vert n_{+}\right\rangle $ and $\left\vert n_{-}\right\rangle
$, which are given by%
\begin{align}
\left\vert n_{+}\right\rangle  &  =\cos\frac{\theta_{n}}{2}\left\vert
n,+\right\rangle +\sin\frac{\theta_{n}}{2}\left\vert n+1,-\right\rangle
\;,\label{|n+>}\\
\left\vert n_{-}\right\rangle  &  =-\sin\frac{\theta_{n}}{2}\left\vert
n,+\right\rangle +\cos\frac{\theta_{n}}{2}\left\vert n+1,-\right\rangle \;,
\label{|n->}%
\end{align}
are eigenstates of $\mathcal{H}_{\mathrm{JC}}$
\cite{Boissonneault_060305,Boissonneault_100504} and the following holds%
\begin{equation}
\mathcal{H}_{\mathrm{JC}}\left\vert n_{\pm}\right\rangle =E_{n_{\pm}%
}\left\vert n_{\pm}\right\rangle \;,
\end{equation}
where%
\begin{equation}
E_{n_{\pm}}=\hbar\left[  \omega_{\mathrm{c}}\left(  n+1\right)  \pm
\frac{\omega_{n}}{2}\right]  \;, \label{E_npm}%
\end{equation}
and where%
\begin{align}
\omega_{n}  &  =\sqrt{\Delta^{2}+4g_{1}^{2}\left(  n+1\right)  }%
\;,\label{omega_n}\\
\Delta &  =\omega_{\mathrm{c}}-\omega_{\mathrm{a}}\;,\label{Delta=}\\
\tan\theta_{n}  &  =-\frac{2g_{1}\sqrt{n+1}}{\Delta}\;. \label{tan(theta_n)}%
\end{align}
The ground state $\left\vert 0,-\right\rangle $ satisfies the relation%
\begin{equation}
\mathcal{H}_{\mathrm{JC}}\left\vert 0,-\right\rangle =E_{\mathrm{g}}\left\vert
0,-\right\rangle \;, \label{H|0,->=}%
\end{equation}
where%
\begin{equation}
E_{\mathrm{g}}=\frac{\hbar\Delta}{2}\; \label{E_g}%
\end{equation}
is the ground state energy.

While in the RWA the term $\mathcal{V}_{\mathrm{BS}}$ is disregarded, its
effect, which gives rise to a Bloch-Siegert shift \cite{Forn_237001}, is
estimated below using perturbation theory. As can be seen from Eq.
(\ref{V_BS}), the perturbation $\mathcal{V}_{\mathrm{BS}}$ is proportional to
$g_{1}$. All diagonal matrix elements of $\mathcal{V}_{\mathrm{BS}}$ in the
basis of eigenstates of $\mathcal{H}_{\mathrm{JC}}$ [see Eqs. (\ref{|n+>}),
(\ref{|n->}) and (\ref{H|0,->=})] vanish, and consequently the lowest
nonvanishing order of the perturbation expansion is the second one. To second
order in $g_{1}$ the energy of the ground state is found to be given by [see
Eqs. (\ref{omega_n}), (\ref{Delta=}) and (\ref{E_g})]%
\begin{equation}
\hbar^{-1}E_{\mathrm{g}}=\frac{\Delta}{2}+\omega_{\mathrm{BS},0}\;,
\label{E_g BS}%
\end{equation}
and the energies of the excited states by%
\begin{align}
\hbar^{-1}E_{n_{\pm}}  &  =\left(  n+1\right)  \left(  \omega_{\mathrm{c}}%
\pm\omega_{\mathrm{BS}}\right) \nonumber\\
&  \pm\sqrt{\frac{\Delta^{2}}{4}+\left(  n+1\right)  g_{1}^{2}}+\omega
_{\mathrm{BS},0}\;,\nonumber\\
&  \label{E_npm BS}%
\end{align}
where%
\begin{equation}
\omega_{\mathrm{BS}}=\frac{g_{1}^{2}}{\omega_{\mathrm{c}}+\omega_{\mathrm{a}}%
}\;, \label{omega_BS}%
\end{equation}
and where%
\begin{equation}
\omega_{\mathrm{BS},0}=-g_{1}^{2}\left(  \frac{1}{\omega_{\mathrm{c}}%
+\omega_{\mathrm{a}}}+\frac{\cot^{2}\theta}{\omega_{\mathrm{c}}}\right)  \;.
\end{equation}
The following holds%
\begin{equation}
\hbar^{-1}\left(  E_{n-}-E_{\mathrm{g}}\right)  =\left(  n+1\right)  \left(
\omega_{\mathrm{c}}-\omega_{\mathrm{BS}}+\frac{g_{1}^{2}}{\Delta}\right)
+O\left(  g_{1}^{4}\right)  \;,
\end{equation}
and%
\begin{equation}
\hbar^{-1}\left(  E_{n+}-E_{0+}\right)  =n\left(  \omega_{\mathrm{c}}%
+\omega_{\mathrm{BS}}-\frac{g_{1}^{2}}{\Delta}\right)  +O\left(  g_{1}%
^{4}\right)  \;,
\end{equation}
thus in the linear regime and when $g_{1}^{2}/\left\vert \Delta\right\vert
\ll1$ the system has two resonance frequencies given by $\omega_{\mathrm{c}%
}\pm\omega_{\mathrm{BS}}\mp g_{1}^{2}/\Delta$.

\section{Superharmonic Resonances}

Superharmonic resonances occur near the points at which the externally applied
flux is tuned such that the ratio $\omega_{\mathrm{a}}/\omega_{\mathrm{c}}$
between the qubit and cavity mode resonance frequencies becomes an integer. In
the analysis below only the averaged system's response is evaluated, and thus
the equations of motion can be simplified by replacing noise terms by their
thermal average, and treating the operators $A$, $\Sigma_{z}$ and $\Sigma_{+}$
as complex numbers, which are labeled by $\alpha_{\mathrm{p}}e^{-i\omega
_{\mathrm{p}}t}$, $P_{z}$ and $P_{+}$, respectively. In this approach Eqs.
(\ref{dA/dt V2}), (\ref{d Sigma_z/dt V2}) and (\ref{d Sigma_+/dt V2}) become
[see Eqs. (\ref{<V_z>=}) and (\ref{<V_+>=})]%
\begin{align}
&  \frac{\mathrm{d}\alpha_{\mathrm{R}}}{\mathrm{d}t}+\left[  -i\Delta
_{\mathrm{pc}}+\gamma_{\mathrm{c}}+\left(  iK_{\mathrm{c}}+\gamma
_{\mathrm{c}4}\right)  \left\vert \alpha_{\mathrm{R}}\right\vert ^{2}\right]
\alpha_{\mathrm{R}}\nonumber\\
&  =ig_{1}\left[  \cot\theta P_{z}-\left(  P_{+}+P_{+}^{\ast}\right)  \right]
e^{i\omega_{\mathrm{p}}t}\nonumber\\
&  -i\sqrt{2\gamma_{\mathrm{c}1}}e^{i\phi_{\mathrm{c}1}}b_{\mathrm{c}%
1}^{\mathrm{in}}\;,\nonumber\\
&  \label{d alpha_R/dt SHR}%
\end{align}%
\begin{equation}
\frac{\mathrm{d}P_{z}}{\mathrm{d}t}+\frac{P_{z}-P_{0}}{T_{1}}+2i\omega
_{\mathrm{g}}\left(  P_{+}-P_{+}^{\ast}\right)  =0\;, \label{dP_z/dt SHR}%
\end{equation}
and%
\begin{equation}
\frac{\mathrm{d}P_{+}}{\mathrm{d}t}-i\omega_{\mathrm{a}}P_{+}+\frac{P_{+}%
}{T_{2}}+i\omega_{\mathrm{g}}\left(  2\cot\theta P_{+}+P_{z}\right)  =0\;,
\label{dP_+/dt SHR}%
\end{equation}
where%
\begin{equation}
\omega_{\mathrm{g}}=g_{1}\left(  \alpha_{\mathrm{R}}e^{-i\omega_{\mathrm{p}}%
t}+\alpha_{\mathrm{R}}^{\ast}e^{i\omega_{\mathrm{p}}t}\right)  \;,
\end{equation}
and where [see Eq. (\ref{tan theta})]%
\begin{equation}
\cot\theta=\frac{\omega_{\mathrm{f}}}{\omega_{\Delta}}\ .
\end{equation}

By employing the transformation%
\begin{equation}
P_{+}=e^{-i\theta_{\mathrm{d}}}P_{\mathrm{d}+}\ ,
\end{equation}
where%
\[
\theta_{\mathrm{d}}=%
{\displaystyle\int\nolimits^{t}}
\mathrm{d}t^{\prime}\;\left(  2\cot\theta\omega_{\mathrm{g}}\left(  t^{\prime
}\right)  -\Delta_{n}-\omega_{\mathrm{a}}\right)  \;,
\]
and where $\Delta_{n}$ is a real constant (to be determined later), Eqs.
(\ref{d alpha_R/dt SHR}), (\ref{dP_z/dt SHR}) and (\ref{dP_+/dt SHR}) become%
\begin{align}
&  \frac{\mathrm{d}\alpha_{\mathrm{R}}}{\mathrm{d}t}+\left[  -i\Delta
_{\mathrm{pc}}+\gamma_{\mathrm{c}}+\left(  iK_{\mathrm{c}}+\gamma
_{\mathrm{c}4}\right)  \left\vert \alpha_{\mathrm{R}}\right\vert ^{2}\right]
\alpha_{\mathrm{R}}\nonumber\\
&  =ig_{1}\left[  \cot\theta P_{z}-\left(  e^{-i\theta_{\mathrm{d}}%
}P_{\mathrm{d}+}+e^{i\theta_{\mathrm{d}}}P_{\mathrm{d}+}^{\ast}\right)
\right]  e^{i\omega_{\mathrm{p}}t}\nonumber\\
&  -i\sqrt{2\gamma_{\mathrm{c}1}}e^{i\phi_{\mathrm{c}1}}b_{\mathrm{c}%
1}^{\mathrm{in}}\;,\nonumber\\
&  \label{d alpha_R/dt SHR V2}%
\end{align}%
\begin{equation}
\frac{\mathrm{d}P_{z}}{\mathrm{d}t}+\frac{P_{z}-P_{0}}{T_{1}}+2i\left(
\zeta_{\mathrm{g}}P_{\mathrm{d}+}-\zeta_{\mathrm{g}}^{\ast}P_{\mathrm{d}%
+}^{\ast}\right)  =0\;, \label{dP_z/dt SHR V2}%
\end{equation}
and%
\begin{equation}
\frac{\mathrm{d}P_{\mathrm{d}+}}{\mathrm{d}t}+\frac{P_{\mathrm{d}+}}{T_{2}%
}+i\Delta_{n}P_{\mathrm{d}+}+i\zeta_{\mathrm{g}}^{\ast}P_{z}=0\;,
\label{dP_+/dt SHR V2}%
\end{equation}
where%
\begin{equation}
\zeta_{\mathrm{g}}=\omega_{\mathrm{g}}e^{-i\theta_{\mathrm{d}}}\;.
\end{equation}

By employing the Jacobi-Anger expansion, which is given by%
\begin{equation}
\exp\left(  iz\cos\varphi\right)  =\sum\limits_{l=-\infty}^{\infty}i^{l}%
J_{l}\left(  z\right)  e^{il\varphi}\;, \label{Jacobi-Anger}%
\end{equation}
where $J_{l}\left(  z\right)  $ is the $l$'th Bessel function of the first
kind, one finds that%
\begin{equation}
e^{-i\theta_{\mathrm{d}}}=\sum\limits_{l=-\infty}^{\infty}\left(
-\frac{\alpha_{\mathrm{R}}^{\ast}}{\left\vert \alpha_{\mathrm{R}}\right\vert
}\right)  ^{l}J_{l}\left(  \frac{4g_{1}\omega_{\mathrm{f}}\left\vert
\alpha_{\mathrm{R}}\right\vert }{\omega_{\mathrm{p}}\omega_{\Delta}}\right)
e^{i\left(  l\omega_{\mathrm{p}}+\Delta_{n}+\omega_{\mathrm{a}}\right)  t}\;.
\end{equation}
Near the $n$'th superharmonic resonance, i.e. when $\omega_{\mathrm{a}}\simeq
n\omega_{\mathrm{p}}$, where $n$ is an integer, the dominant term in the
Jacobi-Anger expansion is the $l^{\prime}$'th one, where $l^{\prime}=1-n$. By
disregarding all other terms in the expansion, choosing the detuning frequency
$\Delta_{n}$ to be given by%
\begin{equation}
\Delta_{n}=n\omega_{\mathrm{p}}-\omega_{\mathrm{a}}\;,
\end{equation}
and disregarding all rapidly oscillating terms, the equations of motion
(\ref{d alpha_R/dt SHR V2}), (\ref{dP_z/dt SHR V2}) and (\ref{dP_+/dt SHR V2})
become%
\begin{align}
&  \frac{\mathrm{d}\alpha_{\mathrm{R}}}{\mathrm{d}t}+\left[  -i\Delta
_{\mathrm{pc}}+\gamma_{\mathrm{c}}+\left(  iK_{\mathrm{c}}+\gamma
_{\mathrm{c}4}\right)  \left\vert \alpha_{\mathrm{R}}\right\vert ^{2}\right]
\alpha_{\mathrm{R}}\nonumber\\
&  =-\frac{i}{\alpha_{\mathrm{R}}^{\ast}}\zeta_{\mathrm{g},n}^{\ast
}P_{\mathrm{d}+}^{\ast}-i\sqrt{2\gamma_{\mathrm{c}1}}e^{i\phi_{\mathrm{c}1}%
}b_{\mathrm{c}1}^{\mathrm{in}}\;,\nonumber\\
&  \label{d alpha_R/dt SHR V3}%
\end{align}%
\begin{equation}
\frac{\mathrm{d}P_{z}}{\mathrm{d}t}+\frac{P_{z}-P_{0}}{T_{1}}+2i\left(
\zeta_{\mathrm{g},n}P_{\mathrm{d}+}-\zeta_{\mathrm{g},n}^{\ast}P_{\mathrm{d}%
+}^{\ast}\right)  =0\;, \label{dP_z/dt SHR V3}%
\end{equation}
and%
\begin{equation}
\frac{\mathrm{d}P_{\mathrm{d}+}}{\mathrm{d}t}+\frac{P_{\mathrm{d}+}}{T_{2}%
}+i\Delta_{n}P_{\mathrm{d}+}+i\zeta_{\mathrm{g},n}^{\ast}P_{z}=0\;,
\label{dP_+/dt SHR V3}%
\end{equation}
where%
\begin{equation}
\zeta_{\mathrm{g},n}=\alpha_{\mathrm{R}}\left(  -\frac{\alpha_{\mathrm{R}%
}^{\ast}}{\left\vert \alpha_{\mathrm{R}}\right\vert }\right)  ^{1-n}g_{n}\;,
\end{equation}
and where%
\begin{equation}
g_{n}=g_{1}J_{1-n}\left(  \frac{4g_{1}\omega_{\mathrm{f}}\left\vert
\alpha_{\mathrm{R}}\right\vert }{\omega_{\mathrm{p}}\omega_{\Delta}}\right)
\; \label{g_n}%
\end{equation}
is the effective coupling coefficient of the $n$'th superharmonic resonance.

At fixed points of the equations of motion the following holds [see Eqs.
(\ref{dP_z/dt SHR V3}) and (\ref{dP_+/dt SHR V3})]%
\begin{equation}
P_{\mathrm{d}+}=-\frac{i\zeta_{\mathrm{g},n}^{\ast}T_{2}P_{z}}{1+i\Delta
_{n}T_{2}}\;,
\end{equation}%
\begin{equation}
P_{0}=\left(  1+\frac{4T_{1}T_{2}\left\vert \zeta_{\mathrm{g},n}\right\vert
^{2}}{1+\Delta_{n}^{2}T_{2}^{2}}\right)  P_{z}\;,
\end{equation}
and thus%
\begin{equation}
P_{\mathrm{d}+}=-\frac{iT_{2}\zeta_{\mathrm{g},n}^{\ast}\left(  1-i\Delta
_{n}T_{2}\right)  P_{0}}{1+\Delta_{n}^{2}T_{2}^{2}+4\left\vert \zeta
_{\mathrm{g},n}\right\vert ^{2}T_{1}T_{2}}\;.
\end{equation}
Substituting into Eq. (\ref{d alpha_R/dt SHR V3}) yields%
\begin{align}
0  &  =\left[  -i\Delta_{\mathrm{pc}}+\gamma_{\mathrm{c}}+\left(
iK_{\mathrm{c}}+\gamma_{\mathrm{c}4}\right)  E_{\mathrm{c}}+i\Upsilon
_{\mathrm{ba},n}P_{0}\right]  \alpha_{\mathrm{R}}\nonumber\\
&  +i\sqrt{2\gamma_{\mathrm{c}1}}e^{i\phi_{\mathrm{c}1}}b_{\mathrm{c}%
1}^{\mathrm{in}}\;,\nonumber\\
&  \label{eq for alpha_R SHR}%
\end{align}
where $E_{\mathrm{c}}=\left\vert \alpha_{\mathrm{R}}\right\vert ^{2}$ and
where%
\begin{equation}
\Upsilon_{\mathrm{ba},n}=\frac{g_{n}^{2}T_{2}\left(  i-\Delta_{n}T_{2}\right)
}{1+\Delta_{n}^{2}T_{2}^{2}+4g_{n}^{2}T_{1}T_{2}E_{\mathrm{c}}}\;.
\label{Upsilon_ba,n}%
\end{equation}
As can be seen by comparing Eqs. (\ref{Upsilon_ba,n}) and (\ref{Upsilon_ba}),
the effect of the qubit on the steady state response of the cavity mode near
the $n^{\prime}$th superharmonic resonance can be taken into account in the
same way as for the case of the primary resonance, provided that $g_{1}$ is
substituted by $g_{n}$ and $\Delta_{1}$ is substituted by $\Delta_{n}$.

\bibliographystyle{apsrev}
\bibliography{C:/Users/eyal/software/swp40/TCITeX/BibTeX/bib/Eyal_bib/Eyal_Bib}